\documentclass{article}
\usepackage{a4wide}

\usepackage{graphicx}%
\usepackage{multirow}%
\usepackage{amsmath,amssymb,amsfonts}%
\usepackage{amsthm}%
\usepackage{mathrsfs}%
\usepackage[title]{appendix}%
\usepackage{xcolor}%
\usepackage{textcomp}%
\usepackage{manyfoot}%
\usepackage{booktabs}%
\usepackage{algorithm}%
\usepackage{algorithmic}%
\usepackage{listings}%

\usepackage{hyperref}       
\usepackage{natbib}
\usepackage{subcaption} 

\newtheorem{definition}{Definition}
\newtheorem{denotation}{Denotation}
\newtheorem{theorem}{Theorem}
\newtheorem{lemma}{Lemma}

\title{Achieving Collective Welfare in Multi-Agent Reinforcement Learning via Suggestion Sharing}

\author{Yue Jin$^\mathbf{1}$, Shuangqing Wei$^\mathbf{2}$, Giovanni Montana$^\mathbf{1,3,4}$\\ 
$^1$Warwick Manufacturing Group, University of Warwick, Coventry, UK\\ 
$^2$School of Electrical Engineering and Computer Science, Louisiana State University, USA\\ 
$^3$Department of Statistics, University of Warwick, Coventry, UK\\ 
$^4$Alan Turing Institute, London, UK\\ 
\texttt{\{yue.jin.3, g.montana\}@warwick.ac.uk, swei@lsu.edu}}

\begin{document}
\maketitle

\abstract{In human society, the conflict between self-interest and collective well-being often obstructs efforts to achieve shared welfare. Related concepts like the Tragedy of the Commons and Social Dilemmas frequently manifest in our daily lives. As artificial agents increasingly serve as autonomous proxies for humans, we propose a novel multi-agent reinforcement learning (MARL) method to address this issue—learning policies to maximise collective returns even when individual agents' interests conflict with the collective one. Unlike traditional cooperative MARL solutions that involve sharing rewards, values, and policies or designing intrinsic rewards to encourage agents to learn collectively optimal policies, we propose a novel MARL approach where agents exchange action suggestions. Our method reveals less private information compared to sharing rewards, values, or policies, while enabling effective cooperation without the need to design intrinsic rewards. Our algorithm is supported by our theoretical analysis that establishes a bound on the discrepancy between collective and individual objectives, demonstrating how sharing suggestions can align agents' behaviours with the collective objective. Experimental results demonstrate that our algorithm performs competitively with baselines that rely on value or policy sharing or intrinsic rewards.  }

\section{Introduction}

Multi-agent reinforcement learning (MARL) enables collaborative decision-making in diverse real-world applications, such as autonomous vehicle control \citep{Xia2022Multi-AgentTracking, Qiu2023ImprovingControl, Jin2021HierarchicalControl}, robotics \citep{Wang2022DistributedReview, Peng2021FACMAC:Gradients, Sun2020ScalingGraph}, and communications systems \citep{SiedlerDynamicReforestation, Huang2022Importance-AwareLearning}.  In these scenarios, artificial agents often act as autonomous decision makers. MARL provides a powerful framework for these settings, enabling agents to learn coordination strategies based on rewards reflecting a common goal.

However, in many cases, a fundamental challenge arises when agents, reflecting the preferences of individuals, are incentivised by interests that conflict with the collective good. This tension is exemplified by the Tragedy of the Commons \citep{Ostrom1990GoverningAction} and Social Dilemmas \citep{Kollock1998SOCIALCooperation, VanLange2013TheReview}, where pursuit of individual interests can lead to collectively harmful outcomes. For instance, when individuals can benefit from a shared resource without contributing to its maintenance, they often face incentives to `free-ride' on others' efforts rather than contribute fairly. Without mechanisms to align individual actions with collective welfare, such systems can collapse into inefficient equilibria where shared resources are depleted or congested, harming all participants. Decades of research in economics and sociology have shown that resolving these dilemmas requires careful mechanism design to foster coordination while respecting individual interests \citep{Hauser2019SocialUnequals, MacyLearningDilemmas, Gersani2001TragedyCompetition, Milinski2002ReputationCommons}.

To illustrate these challenges, consider a smart grid system where consumers balance electricity costs against personal comfort. Each consumer optimises their own trade-off, but electricity costs depend on the collective demand patterns across all users. High simultaneous usage drives up prices for everyone, suggesting that consumers should coordinate to avoid peak times. However, individuals may be reluctant to compromise their comfort, instead hoping others will reduce their consumption. This misalignment between individual comfort optimisation and collective cost minimisation can result in inefficient peak loads and higher costs for all participants. A similar dynamic occurs in traffic networks, where drivers independently choose routes to minimise their personal travel times. Without coordination, too many drivers selecting the same optimal routes create congestion, leading to increased delays for everyone.

A straightforward way to formalise the problem as a MARL problem for collective welfare is to train agents' policies that maximise long-term collective return. Existing solutions often involve introducing designed intrinsic rewards and exchanging individual rewards, values or model parameters. Previous works have proposed various intrinsic rewards based on factors such as social influence, morality, and inequity aversion \citep{Tennant2023ModelingLearning, Hughes2018InequityDilemmas, Jaques2019SocialLearning}. While intrinsic rewards can encourage agents to cooperate, designing appropriate rewards can be intractable in some scenarios.

Alternatively, sharing rewards has been explored as a means to guide agents towards a collective optimum \citep{Chu2020Multi-AgentControl,Yi2022LearningLearning, Chu2020Multi-agentControlb}. Other approaches involve sharing model parameters or the output values of value functions \citep{Zhang2018NetworkedSpaces, Zhang2018FullyAgents, Zhang2020Finite-sampleData, Suttle2020ALearning, Du2022ScalableSystems}. By aggregating individual values or model parameters from neighbouring agents, these methods enable agents to estimate a global value and adjust their policies to maximise it. Similarly, strategies that share policy model parameters rather than value estimates have been proposed \citep{Zhang2019DistributedConsensus, Stankovic2022DistributedWeightings, Stankovic2022ConvergentDifference}, where agents learn a shared joint policy through parameter-sharing and consensus techniques.  While these methods have shown promise in maximising collective returns in some cases, they rely on the assumption that agents can freely exchange potentially sensitive information. Moreover, they may suffer from exploration issues: when cooperation experiences are rare, agents often lack sufficient motivation to cooperate. 

In practice, agents typically do not have access to others' rewards, value functions, or policy functions. For instance, in a smart grid system, consumers' electricity usage policies reflect sensitive information about their daily routines and financial constraints, and their interests (rewards/ values) related to comfort are also private. This information may not be something they are willing to share with other participants or a central coordinator. Similarly, in a traffic network, drivers' routing preferences and time valuations, which  reveal sensitive details about their destinations, schedule constraints, and willingness to pay for faster travel, are rarely shared with others. Traditional MARL approaches that rely on agents sharing rewards, policies, or value estimates thus become problematic in such settings.

Based on these observations, we propose Suggestion Sharing (SS), a novel approach for cooperative policy learning that facilitates effective coordination for collective welfare. SS is grounded in the premise that each agent benefits more when others cooperate, regardless of its own decision to cooperate. For example, in the smart grid scenario, whether or not an agent reduces its electricity usage (cooperates), it always receives a higher reward if other agents cooperate by using less electricity. Thus, agents can share suggestions to encourage cooperation, even in the absence of prior cooperation examples. In SS, agents learn suggestions, share them with one another, and incorporate them into each agent's policy optimisation objective, which is derived from a lower bound of the original collective objective.

Consequently, in SS, instead of sharing policies or rewards, agents exchange only action suggestions—proposals for how others could act to help achieve collective benefits. This iterative process aligns individual behaviours with collective objectives while revealing significantly less private information compared to existing approaches. Empirical results across multiple domains, including sequential social dilemmas and the tragedy of the commons, demonstrate that SS achieves cooperation performance competitive with traditional MARL methods that rely on sharing policies or value functions.

The main contributions of this paper are as follows. We propose a novel Suggestion-Sharing-based MARL (SS)  method to learn cooperative policies for collective welfare when individual interests may conflict with collective objectives. Our method reveals less private information than the traditional cooperative MARL methods that resort to sharing rewards, values, or policies, while enabling effective cooperation without the need to design intrinsic rewards. Theoretically, we show that the optimisation objective of SS serves as a lower bound for the original collective objective. Empirical results demonstrate that SS performs competitively with existing MARL algorithms that rely on sharing policies or values. 

The remainder of this paper is structured as follows. Section \ref{sec:related_work} reviews related work on cooperative MARL under individual reward settings. Section \ref{sec:preliminaties} provides the technical background and problem formulation. Section \ref{sec:methodology} details our methodology, including theoretical foundations and the proposed algorithm. Section \ref{sec:results} outlines the experimental setup and results. Finally, Section \ref{sec:conclusions} discusses the implications of our findings and suggest directions for future research. 


\section{Related Work} \label{sec:related_work}

In this work, we focus on cooperative MARL under individual reward, which is distinguished from numerous contemporary studies that focus on optimising multi-agent policies under the assumption of an evenly split shared team reward \citep{Kuba2022TrustLearning, Wu2021CoordinatedOptimization, Sun2022TrustNon-stationarity,Jiang2022I2QAlgorithm}. Cooperation under individual rewards reflects a more realistic scenario in many real-world applications, where agents need to learn to cooperate based on limited and individual information due to privacy or scalability concerns.

With an individual reward setup, many works \citep{Lowe2017Multi-agentEnvironments, Iqbal2019Actor-attention-criticLearning, Foerster2017StabilisingLearning, Omidshafiei2017DeepObservability, Kim2021CommunicationSharing, Jaques2019SocialLearning} focus on solving Nash equilibrium of a Markov game, i.e., agent seeks the policy that maximises its own expected return. However, that may not result in collective optimum when agents have conflicting individual interests, such as in social dilemmas, which can hinder collective cooperation. Our research focuses on maximising the total return across all agents where each agent needs to cooperate to achieve collective optimum. In the rest of this section, we introduce related works aiming to solve this problem.

\textbf{MARL for Social dilemmas} Social dilemmas highlight the tension between individual pursuits and collective outcomes. In these scenarios, agents aiming for personal gains can lead to compromised group results. For instance, one study has explored self-driven learners in sequential social dilemmas using independent deep Q-learning \citep{Leibo2017Multi-agentDilemmas}. A prevalent research direction introduces intrinsic rewards to encourage collective-focused policies. For example, \textit{moral learners} have been introduced with varying intrinsic rewards \citep{Tennant2023ModelingLearning} while other approaches have adopted an inequity-aversion-based intrinsic reward \citep{Hughes2018InequityDilemmas} or rewards accounting for social influences and predicting other agents' actions \citep{Jaques2019SocialLearning}. Borrowing from economics, our method integrated formal contracting to motivate global collaboration \citep{Christoffersen2023GetRL}. While these methods modify foundational rewards, we maintain original rewards, emphasizing a collaborative, information-sharing strategy to nurture cooperative agents.

\textbf{Value sharing} Value sharing methods use shared Q-values or state-values among agents to better align individual and collective goals. Many of these methods utilize consensus techniques to estimate the value of a joint policy and guide individual policy updates accordingly. For instance, a number of networked actor-critic algorithms exist based on value function consensus, wherein agents merge individual value functions towards a global consensus by sharing parameters \citep{Zhang2018NetworkedSpaces, Zhang2018FullyAgents,  Zhang2020Finite-sampleData, Suttle2020ALearning}. Instead of sharing value function parameters, \citep{Du2022ScalableSystems} shares function values for global value estimation. However, these methods have an inherent limitation: agents modify policies individually using fixed Q-values or state-values, making them less adaptive to immediate policy shifts from peers and potentially introducing policy discoordination. In contrast, our approach enables more adaptive coordination by having agents directly share and respond to peer suggestions.

\textbf{Reward sharing} Reward sharing is about receiving feedback from a broader system-wise outcome perspective, ensuring that agents act in the collective best interest of the group. Some works have introduced a spatially discounted reward function \citep{Chu2020Multi-AgentControl, Chu2020Multi-agentControlb}. In these approaches, each agent collaboratively shares rewards within its vicinity. Subsequently, an adjusted reward is derived by amalgamating the rewards of proximate agents, with distance-based discounted weights. Other methods advocate for the dynamic learning of weights integral to reward sharing, which concurrently evolve as agents refine their policies \citep{Yi2022LearningLearning}. In our research, we focus on scenarios where agents know only their individual rewards and are unaware of their peers' rewards. This mirrors real-world situations where rewards are kept confidential or sharing rewards suffers challenges such as communication delays and errors. Consequently, traditional value or reward sharing methods fall short in these contexts. In contrast, our method induces coordination without requiring reward sharing.

\textbf{Policy sharing} Policy sharing strives to unify agents' behaviors through an approximate joint policy. However, crafting a global policy for each agent based on its individual reward can lead to suboptimal outcomes. Consensus update methods offer a solution by merging individually learned joint policies towards an optimal joint policy. Several studies have employed such a strategy, focusing on a weighted sum of neighboring agents' policy model parameters \citep{Zhang2019DistributedConsensus, Stankovic2022DistributedWeightings, Stankovic2022ConvergentDifference}. These methods are particularly useful when sharing individual rewards or value estimates is impractical. Yet, sharing policy model parameters risks added communication overheads and data privacy breaches. PS is based on the idea of federated learning and shares the parameters of joint policies among agents. In contrast, our method focuses on learning individual policies and sharing only the relevant action distributions of the suggesting policies with the corresponding agents, which typically involves less communication overhead compared to sharing entire policy parameters with all the neighbouring agents.

\textbf{Teammate modeling} Teammate/opponent modeling in MARL often relies on agents having access to, or inferring, information about teammates' goals, actions, or rewards. This information is then used to improve collective outcomes \citep{Albrecht2018AutonomousProblems, He2016, Wen2019, Zheng2018}. Our approach differs from traditional team modeling. Rather than focusing on predicting teammates' exact actions or strategies, our method has each agent calculate and share action suggestions that would benefit its own strategy. These suggestions are used by other agents (not the agent itself) to balance their objectives with those of the agent sending the suggestion. This approach emphasizes suggestions that serve the agent's own objective optimisation. Coordination occurs through policy adaptation based on \emph{others' suggestions} that implicitly include information about their returns, rather than modeling their behaviors. It contrasts with conventional team modeling in MARL that focuses on modeling teammates' behaviors directly.


\section{Preliminaries and Problem Statement} \label{sec:preliminaties}

To optimise the collective welfare, we formulate the problem as a Multi-agent Markov Decision Process (MMDP). Specifically, we consider an MMDP with $N$ agents represented as a tuple $<\mathcal{S}, \{\mathcal{A}^i\}_{i=1}^N, \mathcal{P}, \{\mathcal{R}^i\}_{i=1}^N, \gamma>$, where $\mathcal{S}$ denotes a global state space, $\mathcal{A}^i$ is the individual action space, $\mathcal{A} = \Pi_{i=1}^N\mathcal{A}^i$ is the joint action space, $\mathcal{P}: \mathcal{S} \times \mathcal{A} \times \mathcal{S} \rightarrow [0,1]$ is the state transition function, $\mathcal{R}^i: \mathcal{S} \times \mathcal{A} \rightarrow \mathbb{R}$ is the individual reward function, and $\gamma$ is a discount factor.
Each agent $i$ selects an action $a^i \in \mathcal{A}^i$ based on its individual policy $\pi^i: \mathcal{S} \times \mathcal{A}^i \rightarrow [0, 1]$.  
The joint action of all agents is represented by $\boldsymbol{a} \in \mathcal{A}$, and the joint policy across these agents is denoted as $\boldsymbol{\pi}(\cdot|s) = \prod_{i=1}^N \pi^i(\cdot|s)$. The objective is to maximise the expectation of collective cumulative return of all agents,
\begin{equation} \label{objective}
  \eta(\boldsymbol{\pi})= \sum_{i=1}^N \mathbb{E}_{\tau \sim \boldsymbol{\pi}}\left[\sum_{t=0}^\infty \gamma^t r_t^i\right], 
\end{equation}
where the expectation, $\mathbb{E}_{\tau \sim \boldsymbol{\pi}}[\cdot]$, is computed over trajectories with an initial state distribution $s_0 \sim d(s_0)$, action selection $\boldsymbol{a}_t \sim \boldsymbol{\pi}(\cdot| s_t)$, state transitions $s_{t+1} \sim \mathcal{P}(\cdot| s_t, \boldsymbol{a}_t)$, and 
$r_t^i = \mathcal{R}^i(s, \boldsymbol{a})$ is the reward for individual agent $i$. Here, we use $r^i_t = R^i(s, a)$ for simplicity of notation, but this can be easily extended to a stochastic reward function without affecting the core of our method.
An individual advantage function is defined as: 
\begin{equation}
A_i^{\boldsymbol{\pi}}(s, \boldsymbol{a}) = Q_i^{\boldsymbol{\pi}} (s, \boldsymbol{a}) - V_i^{\boldsymbol{\pi}} (s)
\end{equation}
which depends on the individual state-value  and  action-value functions, respectively,
\begin{equation} \label{value_function}
\begin{aligned}
    &V_i^{\boldsymbol{\pi}} (s) = \mathbb{E}_{\tau \sim \boldsymbol{\pi}}\left[\sum_{t=0}^\infty \gamma^t r_t^i | s_0 = s \right], 
    & Q_i^{\boldsymbol{\pi}} (s, \boldsymbol{a}) = \mathbb{E}_{\tau \sim \boldsymbol{\pi}}\left[\sum_{t=0}^\infty \gamma^t r_t^i | s_0 = s, \boldsymbol{a}_0 = \boldsymbol{a} \right].
\end{aligned}
\end{equation}

MMDP has also been employed in previous works. \citep{Zhao2020DistributedLearning, Krouka2022Communication-EfficientLearning} formalised the same problem as we did. \citep{Chen2022Communication-EfficientLearning} considered a similar problem but included a central controller that collects information from all agents.  \citep{Zhang2018FullyAgents, Du2022ScalableSystems, Sha2021PolicyNetworks} used the same basic problem formalism, but added a network structure on agent systems, referring to it as Networked MMDP or MARL over networks. Additionally, \citep{Lei2022AdaptiveIoT} presented the Networked MARL problem from the perspective of Alternating Direction Method of Multipliers (ADMM).

However, in our setup, agents do not have direct access to others' policies, rewards, or values. This setting is particularly relevant for applications where users prefer not to reveal their exact policies and rewards or values. Our work aims to bridge this gap between individual and collective return maximisation. It enables agents to approximate the optimisation of the collective objective while operating solely with their individual reward signals. 
In the next section, we present a method where agents iteratively share suggestions to maximise a lower bound of Eq.\ref{objective}. This method is general and not dependent on any specific protocol for communicating suggestions between agents. In Sec.\ref{practical_algo}, we propose a practical algorithm that involves sharing information within agents' neighbourhoods. Our experiments demonstrate the effects of different sharing protocols on the performance of MARL cooperation.
For convenience, notations frequently used in this paper are listed in Table~\ref{tab.notation}.

\begin{table}[!t]
\renewcommand\arraystretch{1.2}
\footnotesize
\captionsetup{font=small}
\caption{Notations frequently used in this paper.}
\centering
\begin{tabular}{l|l}
\toprule[1pt]
 $\eta$ & The expectation of collective
cumulative return of all agents\\
 \hline
 $\pi^i$ & Individual policy \\
\hline
 $\boldsymbol{\pi}$  & Joint policy\\
\hline
 $\pi^{ij}$ & Suggestion of agent $i$ about agent $j$'s policy when $j \neq i$ \\
\hline
 $\pi^{ii}$ & Equivalent to $\pi^{i}$, which is agent $i$'s own policy\\
\hline
 $\boldsymbol{\tilde{\pi}}^i$  & Suggesting joint policy of agent $i$\\
\hline
 $\boldsymbol{\tilde{\Pi}}$  & Collection of all suggesting joint policies across agents, i.e., $(\boldsymbol{\tilde{\pi}}^1, \cdots, \boldsymbol{\tilde{\pi}}^i, \cdots, \boldsymbol{\tilde{\pi}}^N)$\\
\hline
 $A_i^{\boldsymbol{\pi}}$ & Individual advantage function under joint policy $\boldsymbol{\pi}$\\
\hline
 $\hat{A}_i$ & Estimated value of individual advantage function  \\
\hline
 $\theta^{ij}$  &  Parameters of $\pi^{ij}$\\
 \hline
 $\mathcal{N}_i$  & Agent $i$'s  neighbours \\
\hline
 $\boldsymbol{\theta}^{-ii}$ & Parameters of all the $\pi^{ij}$ ($j \in \mathcal{N}_i$)\\
\bottomrule[1pt]
\end{tabular} \label{tab.notation}
\end{table}

\section{Methodology} \label{sec:methodology}

In this section, we start from solving Eq. \ref{objective}, the collective optimisation objective formulated in Section 3. We derive a lower bound of this objective based on trust region policy optimisation (TRPO) work \citep{Schulman2015TrustOptimization}. The lower bound applies to the setting where agents have individual rewards, distinguishing from previous works where agents share team rewards \citep{Wu2021CoordinatedOptimization, Su2022DecentralizedOptimization}. Then we introduce Suggesting Policies to replace the other agents' policies in the individual term corresponding to each agent in the lower bound and derive a bound for the gap caused by such a replacement. By leveraging the gap, agents can learn policies to maximise the collective return in an individual way without explicit reward or policy sharing. We will see that the gap for each agent is related with the discrepancy between the action distribution suggested by others and the agent's own action distribution. Practically, we propose SS algorithm, where agents share their action suggestions with each other. These suggestions are then considered by other agents when maximising their individual objectives, enabling each agent to align with the collective goal.

Unlike traditional methods that share explicit rewards or objectives, SS involves agents exchanging suggestions that implicitly contain information about others' objectives. By observing how its actions align with aggregated suggestions, each agent can perceive the divergence between its individual interests and the collective goals. This drives policy updates to reduce the identified discrepancy, bringing local and global objectives into closer alignment.

\subsection{Theoretical Developments}

We commence our technical developments by analysing joint policy shifts based on global information. This extends foundational TRPO to multi-agent settings with individual advantage values. We prove the following bound on the expected return difference between new and old joint policies:

\begin{lemma} \label{theorem_1}
We establish a lower bound for expected collective returns:

\begin{equation}\label{TRPO}
    \begin{aligned}
        \eta(\boldsymbol{\pi}_{new}) \geq \eta(\boldsymbol{\pi}_{old}) + \zeta_{\boldsymbol{\pi}_{old}}(\boldsymbol{\pi}_{new}) - C \cdot D_{KL}^{max}(\boldsymbol{\pi}_{old} || \boldsymbol{\pi}_{new}),
    \end{aligned}
\end{equation}
where 
\begin{equation}
\begin{aligned}
     & \zeta_{\boldsymbol{\pi}_{old}}(\boldsymbol{\pi}_{new}) = \mathbb{E}_{s \sim d^{\boldsymbol{\pi}_{old}}(s), \boldsymbol{a}\sim \boldsymbol{\pi}_{new}( |s)} \left[\sum_{i} A_i^{\boldsymbol{\pi}_{old}}(s, \boldsymbol{a})\right],   \quad C = \frac{4 \max_{s, \boldsymbol{a}}|\sum_{i} A_i^{\boldsymbol{\pi}_{old}}(s, \boldsymbol{a})|\gamma}{(1-\gamma)^2}\\
     & D_{KL}^{max}(\boldsymbol{\pi}_{old} || \boldsymbol{\pi}_{new}) = \max_{s}D_{KL}(\boldsymbol{\pi}_{old}(\cdot|s) || \boldsymbol{\pi}_{new}(\cdot|s)).
\end{aligned}
\end{equation}
\end{lemma}

The proof is given in Appendix~\ref{proof_1}. 

The key insight is that the improvement in returns under the new policy depends on both the total advantages of all the agents, as well as the divergence between joint policy distributions. This quantifies the impact of joint policy changes on overall system performance given global knowledge, extending trust region concepts to multi-agent domains.

However, as the improvement in returns is measured by joint policy distributions and total advantages of all agents, it is hard to be used by single agent in MARL settings where  each agent has no access to others' policies and rewards. To address this limitation, we first introduce the concept of \textit{suggesting joint policy} from each agent's local perspective to replace the true joint policy. As we will show in Sec.~\ref{surrogate}, the suggesting joint policy of each agent is solved by optimising an individual objective. Analysing suggesting policies is crucial for assessing the discrepancy between individual objectives and the collective one in cooperative MARL.

\begin{denotation}
For each agent in a multi-agent system, we denote the \textbf{suggesting joint policy} as $\boldsymbol{\tilde{\pi}}^i$, formulated as $\boldsymbol{\tilde{\pi}}^i(\boldsymbol{a}|s) = \prod_{j=1}^N \pi^{ij}(a^j|s)$. Here, for each agent $i$, $\pi^{ij}$ represents the suggestion of agent $i$ about agent $j$'s policy when $j \neq i$. When $j = i$, we have $\pi^{ii} = \pi^{i}$, which is agent $i$'s own policy. To represent the collection of all such suggesting joint policies across agents, we use the notation $\boldsymbol{\tilde{\Pi}} :=(\boldsymbol{\tilde{\pi}}^1, \cdots, \boldsymbol{\tilde{\pi}}^i, \cdots, \boldsymbol{\tilde{\pi}}^N)$. 
\end{denotation}

The suggesting joint policy represents an agent's perspective of the collective strategy constructed from its own policy and suggestions to peers. We will present how to solve such suggesting joint policy in Sec.~\ref{surrogate}.

\begin{definition} \label{def_zeta}
The total expectation of individual advantages over the suggesting joint policies and a common state distribution, is defined as follows:
\begin{equation}
    \zeta_{\boldsymbol{\pi}'}(\boldsymbol{\tilde{\Pi}})=\sum_{i}\mathbb{E}_{s \sim d^{\boldsymbol{\pi}'}(s), \boldsymbol{a}\sim \boldsymbol{\tilde{\pi}}^i(\boldsymbol{a}|s)} \left[A_i^{\boldsymbol{\pi}'}(s, \boldsymbol{a})\right],
\end{equation}
which represents the sum of expected advantages for each agent $i$, calculated over their suggesting joint policy $\boldsymbol{\tilde{\pi}}^i$ and a shared state distribution, $d^{\boldsymbol{\pi}'}(s)$. The advantage $A_i^{\boldsymbol{\pi}'}(s, \boldsymbol{a})$ for each agent is evaluated under a potential joint policy $\boldsymbol{\pi}'$, which may differ from the true joint policy $\boldsymbol{\pi}$ in play. This definition captures the expected benefit each agent anticipates based on the suggesting joint actions, relative to the potential joint policy $\boldsymbol{\pi}'$.
\end{definition}

This concept quantifies the expected cumulative advantage an agent could hypothetically gain by switching from a reference joint policy to the suggesting joint policies of all agents. It encapsulates the perceived benefit of the suggesting policies versus a collective benchmark. Intuitively, if an agent's suggestions are close to the actual policies of other agents, this expected advantage will closely match the actual gains. However, discrepancies in suggestions will lead to divergences, providing insights into the impacts of imperfect local knowledge.

Equipped with these notions of suggesting joint policies and total advantage expectations, we can analyse the discrepancy of the expectation of the total advantage caused by policy shift from the true joint policy, $\boldsymbol{\pi}$, to the individually suggesting ones, $\boldsymbol{\tilde{\Pi}}$. Specifically, we prove the following bound relating this discrepancy:

\begin{lemma} \label{theorem_2}
The discrepancy between $\zeta_{\boldsymbol{\pi}'}(\boldsymbol{\tilde{\Pi}})$ and $\zeta_{\boldsymbol{\pi}'}(\boldsymbol{\pi})$ is upper bounded as follows:
\begin{equation} \label{bound}
\begin{aligned}
     & \zeta_{\boldsymbol{\pi}'}(\boldsymbol{\tilde{\Pi}}) - \zeta_{\boldsymbol{\pi}'}(\boldsymbol{\pi}) \leq 
     f^{\boldsymbol{\pi}'} + \sum_{i} \frac{1}{2} \max_{s, \boldsymbol{a}} \left| A_i^{\boldsymbol{\pi}'}(s, \boldsymbol{a}) \right| 
      \cdot \sum_{s, \boldsymbol{a}}\left(\boldsymbol{\tilde{\pi}}^i(\boldsymbol{a}|s) - \boldsymbol{\pi}(\boldsymbol{a}|s)\right)^2, 
\end{aligned}
\end{equation}
where 
\begin{equation} \label{f_pi_main}
    f^{\boldsymbol{\pi}'} = \sum_{i} \frac{1}{2} \max_{s, \boldsymbol{a}} \left| A_i^{\boldsymbol{\pi}'}(s, \boldsymbol{a}) \right|  \cdot |\mathcal{A}| \cdot \Vert d^{\boldsymbol{\pi}'} \Vert_2^2,
\end{equation}
\end{lemma}
and $\|d^{\pi'}\|_2^2 := \sum_s (d^{\pi'}(s))^2$.

The proof is given in Appendix~\ref{proof_2}.

This result quantifies the potential drawbacks of relying on imperfect knowledge in cooperative MARL settings, where agents' suggestions may diverge from actual peer policies. It motivates reducing the difference between the suggesting and true joint policies.

Previous results bounded the deviation between total advantage expectations under the true joint policy versus under suggesting joint policies. We now build on this to examine how relying too much on past experiences and suggesting joint policies can lead to misjudging the impact of new joint policy shifts over time. To this end, we consider the relationship between $\zeta_{\boldsymbol{\pi}_{\text{old}}}(\boldsymbol{\tilde{\Pi}}_{\text{new}})$, the perceived benefit of the new suggesting joint policies $\boldsymbol{\tilde{\Pi}}_{\text{new}}$, assessed from the perspective of the previous joint policy $\boldsymbol{\pi}_{\text{old}}$, and  $\eta(\boldsymbol{\pi}_{\text{new}})$, which measures the performance of the new joint policy. Specifically, $\zeta_{\boldsymbol{\pi}_{\text{old}}}(\boldsymbol{\tilde{\Pi}}_{\text{new}})$ is defined like Definition~\ref{def_zeta} as:
    \begin{equation}
    \zeta_{\boldsymbol{\pi}_{old}}(\boldsymbol{\tilde{\Pi}}_{new})=\sum_{i}\mathbb{E}_{s \sim d^{\boldsymbol{\pi}_{old}}(s), \boldsymbol{a}\sim \boldsymbol{\tilde{\pi}}_{new}^i(\boldsymbol{a}|s)} \left[A_i^{\boldsymbol{\pi}_{old}}(s, \boldsymbol{a})\right],
    \end{equation}
which represents a potentially myopic and individual perspective informed by the advantage values, $A_i^{\boldsymbol{\pi}_{old}}$, of past policies, as well as individually suggesting joint policies, $\boldsymbol{\tilde{\pi}}_{new}^i$, and thus, it may inaccurately judge the actual impact of switching to $\boldsymbol{\pi}_{\text{new}}$ as quantified by $\eta(\boldsymbol{\pi}_{\text{new}})$. The following theorem provides a lower bound of the collective return, $\eta(\boldsymbol{\pi}_{\text{new}})$, of the newer joint policy, based on $\zeta_{\boldsymbol{\pi}_{\text{old}}}(\boldsymbol{\tilde{\Pi}}_{\text{new}})$.

\begin{theorem} \label{theorem_surrogate}
    Based on suggesting joint policies, a lower bound of the collective return of the true joint policy is given as:
    \begin{equation} \label{lower_bound}
    \begin{aligned}
        \eta(\boldsymbol{\pi}_{new}) \geq &\eta(\boldsymbol{\pi}_{old}) + \zeta_{\boldsymbol{\pi}_{old}}(\boldsymbol{\tilde{\Pi}}_{new}) - C \cdot \sum_i D_{KL}^{max}(\pi^{ii}_{old} || \pi^{ii}_{new}) -  \\
        & f^{\boldsymbol{\pi}_{old}} - \sum_{i} \frac{1}{2} \max_{s, \boldsymbol{a}} \left| A_i^{\boldsymbol{\pi}_{old}}(s, \boldsymbol{a}) \right|  \cdot \sum_{s, \boldsymbol{a}}\left(\boldsymbol{\tilde{\pi}}^i_{new}(\boldsymbol{a}|s) - \boldsymbol{\pi}_{new}(\boldsymbol{a}|s)\right)^2.
    \end{aligned}
    \end{equation}
\end{theorem}
The full proof is given in Appendix~\ref{proof_3}.  This theorem explains the nuanced dynamics of policy changes in MARL where agents learn separately. It sheds light on how uncoordinated local updates between individual agents affect the collective performance. At the same time, this result suggests a potential way to improve overall performance by leveraging the suggesting joint policies held by each agent.

\subsection{A Surrogate Optimisation Objective} \label{surrogate}

Our preceding results established analytical foundations for assessing joint policy improvement in multi-agent settings with individual advantage values and suggesting joint policies. We now build upon these results to address the practical challenge of optimising collective returns when agents lack knowledge of others' policies, rewards, and values.

Directly maximising the expected collective returns, $\eta(\boldsymbol{\pi})$, is intractable without global knowledge of the joint policy and collective return. However, Theorem \ref{theorem_surrogate} provides insight into a more tractable approach: agents can optimise a localized surrogate objective, $\zeta_{\boldsymbol{\pi}_{\text{old}}}(\boldsymbol{\tilde{\Pi}})$, which is the sum of individual objectives concerning suggesting joint policies and individual advantage values. This simplifies the global objective into an individual form dependent on the suggesting joint policy that is composed of an agent's individual policy, $\pi^{ii}$, and its suggestions for others, $\pi^{ij}$.

To leverage this insight, we use the lower bound given by Theorem \ref{theorem_surrogate}. By maximising this lower bound 
, we can maximise the collective return. We can ignore the terms $\eta(\boldsymbol{\pi}_{\text{old}})$ and $f^{\boldsymbol{\pi}_{\text{old}}}$ from Theorem \ref{theorem_surrogate} in our optimisation problem, as they are not relevant to optimising $\boldsymbol{\tilde{\Pi}}$ and their values are usually bounded. To be specific, the value of $\eta(\boldsymbol{\pi}_{\text{old}})$ is bounded as the reward value is bounded. For $f^{\boldsymbol{\pi}_{\text{old}}}$, as defined in Eq. \ref{f_pi_main}, its value is also bounded since (1) We focus on scenarios with finite and relatively small action spaces (each agent’s discrete action set typically consists of 2–10 actions), which are common in many real-world applications, so $|\mathcal{A}|$ (the size of the action space) is not excessively large. (2) The term $\|d^{\pi_\text{old}}\|^2_2$ is the square L2-norm of the state visitation distribution, which is bounded.(3) The advantage function $A^{\boldsymbol{\pi}_\text{old}}_i(s,a)$ is also bounded as the reward value is bounded.  

Consequently, we propose the following constrained optimisation problem as a surrogate for the original collective objective:
\begin{equation} \label{global_constrained_obj}
\begin{aligned}
    &\max_{\boldsymbol{\tilde{\Pi}}} \sum_{i}\mathbb{E}_{s \sim d^{\boldsymbol{\pi}_{old}}(s), \boldsymbol{a}\sim {\boldsymbol{\tilde{\pi}}^i}(\boldsymbol{a}|s)} \left[A_i^{\boldsymbol{\pi}_{old}}(s, \boldsymbol{a})\right] \\
    & \mathrm{ s.t. } \quad \sum_i D_{KL}^{max}(\pi^{ii}_{old} || \pi^{ii}) \leq \delta,  \qquad ~ 
    \sum_i \max_{s, \boldsymbol{a}} \left| A_i^{\boldsymbol{\pi}_{old}}(s, \boldsymbol{a}) \right|  \cdot \sum_{s, \boldsymbol{a}}  \left(\boldsymbol{\tilde{\pi}}^i(\boldsymbol{a}|s) - \boldsymbol{\pi}(\boldsymbol{a}|s) \right)^2 \leq \delta' .
\end{aligned}
\end{equation}
Note that, taking into account of the results given by \citep{Schulman2015TrustOptimization}, we do not directly include the lower bound of the discrepancy given by Eq.~\ref{lower_bound} in Eq.~\ref{global_constrained_obj}, but instead use constraints to facilitate learning.

Eq.~\ref{global_constrained_obj} captures the essence of coordinating joint policies to maximise individual advantages with suggesting joint policies. However, it still assumes full knowledge of $\boldsymbol{\tilde{\Pi}}$.
To make this feasible in individual policy learning, we reformulate it from each agent's perspective.  Remarkably, we can distill the relevant components into an individual objective and constraints for each individual agent $i$, as follows:
\begin{equation}\label{constrain_obj}
\begin{aligned}
    &\max_{\boldsymbol{\tilde{\pi}}^i} \mathbb{E}_{s \sim d^{\boldsymbol{\pi}_{old}}(s), \boldsymbol{a}\sim \boldsymbol{\tilde{\pi}}^i(\boldsymbol{a}|s)} \left[A_i^{\boldsymbol{\pi}_{old}}(s, \boldsymbol{a}) \right] \\
    & \mathrm{s.t.:} \quad \text{(a)} \quad D_{KL}^{max}(\pi^{ii}_{old} || \pi^{ii}) \leq \delta_1, \quad \text{(b)} \quad \kappa_i \cdot \sum_{s, a_j} (\pi^{ij}(a_j|s) - \pi^{jj}(a_j|s))^2 \leq \delta_2, ~ \forall j \neq i, \\
    & \qquad \quad \text{(c)} \quad \kappa_i \cdot \sum_{s, a_i} (\pi^{ii}(a_i|s) - \pi^{ji}(a_i|s))^2 \leq \delta_2, ~ \forall j \neq i,
\end{aligned}
\end{equation}
where $\kappa_i = \max_{s, \boldsymbol{a}} \left| A_i^{\boldsymbol{\pi}_{old}}(s, \boldsymbol{a}) \right| $.

The constraints in Eq. 12 are imposed on $\pi^{ii}$ and $\pi^{ij}$ ($j \neq i$), which together compose $\boldsymbol{\tilde{\pi}}^i$. Therefore, these constraints effectively limit the space of possible $\boldsymbol{\tilde{\pi}}^i$ by constraining its components. Constraint (a) limits how much the agent's own policy can change, while constraints (b) and (c) ensure that the suggestions are close to the actual policies of other agents. The corresponding terms in these constrains are bounded by some constants or functions, so that they can remain finite. This boundedness aims to guarantee that the discrepancy between the collective and individual objectives is controllable.

The constraints also depend on other agents' policies $\pi^{jj}$ and their suggestions for agent $i$'s policy, $\pi^{ji}$. To enable the evaluation of these terms, each agent $j$ shares its action distribution $\pi^{jj}(\cdot|s)$ and the action distribution suggestion $\pi^{ji}(\cdot|s)$ with agent $i$. This sharing enables each agent $i$ to assess the constraint terms, which couples individual advantage optimisations under local constraints. These constraints reflect both the differences between the policies of others and an agent's suggestions on them, as well as the discrepancy between an agent's own policy and others' suggestions on it. By distributing the optimisation while exchanging policy suggestions, this approach balances individual policy updates while maintaining global coordination among agents.

It's important to distinguish our method from teammate modeling. In teammate modeling, agent $i$ typically approximates peer policies $\hat{\pi}^{ij}$ and uses these approximations when solving for its own policy $\pi^{ii}$. In contrast, our approach in Eq.~\ref{constrain_obj} aims to optimise the suggestions $\pi^{ij}$ alongside $\pi^{ii}$. These optimised suggestions $\pi^{ij}$ are then used by agent $j$ to solve for its policy $\pi^{jj}$. This method allows the suggestions to implicitly incorporate information about individual objectives. Through the exchange of these suggestions, individual agents can balance others' objectives and, consequently, the collective performance while optimising their own objectives.

\subsection{A Practical Algorithm for MARL with SS} \label{practical_algo}

We propose a structured approach to optimise the objective in Eq. \ref{constrain_obj}. The derivation of the algorithm involves specific steps, each targeting different aspects of the optimisation challenge.
Note that in this practical algorithm, we present a setup where agent $i$ exchanges information with neighbours $\{j|j \in \mathcal{N}_i\}$ that may not include all other $(N-1)$ agents, and is not subject to a particular protocol used for determining $\mathcal{N}_i$. 
In experiments, we use different neighbourhood definitions/protocols to investigate corresponding effects.

\subsubsection*{Step 1: Clipping Policy Ratio for KL Constraint}

Addressing the KL divergence constraint (a) in Eq.~\ref{constrain_obj} is crucial in ensuring each agent's policy learning process remains effective. This constraint ensures that updates to an agent's individual policy do not deviate excessively from its previous policy. To manage this, we incorporate a clipping mechanism, inspired by PPO-style clipping \citep{Schulman2017ProximalAlgorithms}, adapted for individual agents in our method.

We start by defining probability ratios for the individual policy and suggesting policies for peers:
\begin{equation}
\xi_i = \frac{\pi^{ii}(a_i|s';\theta^{ii})}{\pi^{ii}_{old}(a_i|s'; \theta^{ii}_{old})}, \quad 
\xi_{\mathcal{N}_i} = \prod_{j \in \mathcal{N}_i}\frac{\pi^{ij}(a_j|s; \theta^{ij})}{\pi^{jj}_{old}(a_j|s; \theta^{jj}_{old})}.
\end{equation}
These ratios measure the extent of change in an agent's policy relative to its previous one and its suggestions to others' true policies. 
We then apply a clipping operation to $\xi_i$, the individual policy ratio:
\begin{equation*}
\begin{aligned}
    &\mathbb{E}_{s \sim d^{\boldsymbol{\pi}_{old}}(s), \boldsymbol{a}\sim \boldsymbol{\pi}_{old}(\boldsymbol{a}|s)} \left[\min \left(\xi_i  \xi_{\mathcal{N}_i} \hat{A}_i, \text{clip}(\xi_i, 1- \epsilon, 1+\epsilon) \xi_{\mathcal{N}_i} \hat{A}_i \right) \right].
\end{aligned}
\end{equation*}
This method selectively restricts major changes to the individual policy $\pi^{ii}$, while allowing more flexibility in updating suggestions on peer policies. It balances the adherence to the KL constraint with the flexibility needed for effective learning and adaptation in a multi-agent environment.

\subsubsection*{Step 2: Penalizing Suggestion Discrepancies}

The objective of this step is to enforce constraints (b) and (c) in Eq. \ref{constrain_obj}, which aim to penalize discrepancies between the suggesting policies and others' policies. Simply optimising the advantage function may not sufficiently increase these discrepancies. To be specific, if $\hat{A}_i > 0$, according to the main objective function, Eq. \ref{constrain_obj}, the gradient used to update $\pi^{ij}$ will be positive and will lead to the increase of $\pi^{ij}$. If $\frac{\pi^{ij}(a|s, \theta^{ij})}{\pi^{jj}(a|s)} < 1$, i.e. $\pi^{ij}(a|s, \theta^{ij}) < \pi^{jj}(a|s)$, then the gradient caused by the main objective will decrease the discrepancy between $\pi^{ij}$ and $\pi^{jj}$.
Therefore, we introduce penalty terms that are activated when policy updates inadvertently increase these discrepancies. Specifically, we define state-action sets $X^{ij}$ to identify where the policy update driven by the advantage exacerbates the discrepancies between the resulting suggesting policies and other agents' current policies, and $X^{ii}$ to identify the discrepancies between the resulting agent's own policy and the ones suggested by other agents. These are defined as: 
\begin{equation}
\begin{aligned}
    X^{ij} &= \left\{(s, \boldsymbol{a}) \mid \frac{\pi^{ij}(a_j|s;\theta^{ij})}{\pi^{jj}(a_j|s)}  \hat{A}_i \geq \hat{A}_i \right\} \qquad
    X^{ii} &= \left\{(s, \boldsymbol{a}) \mid \frac{\pi^{ii}(a_i|s;\theta^{ii})}{\pi^{ji}(a_i|s)}  \hat{A}_i \geq \hat{A}_i \right\},
\end{aligned}
\end{equation}
where the pairs $(s, \boldsymbol{a})$ represent scenarios in which the gradient influenced by $\hat{A}_i$ increases the divergence between the two policies. The following indicator function captures this effect:
\begin{equation}
    \mathbb{I}_X(s, \boldsymbol{a}) = 
    \begin{cases} 
        1 & \text{if } (s, \boldsymbol{a}) \in X, \\
        0 & \text{otherwise}.
    \end{cases}
\end{equation}

\subsubsection*{Step 3: Dual Clipped Objective}

In the final step, we combine the clipped surrogate objective with coordination penalties to form our dual clipped objective:
\begin{equation} \label{obj}
\begin{aligned}
    &\max_{\theta^{ii}, \boldsymbol{\theta}^{-ii}} \mathbb{E}_{s \sim d^{\boldsymbol{\pi}_{old}}(s), \boldsymbol{a}\sim \boldsymbol{\pi}_{old}(\boldsymbol{a}|s)} \left[\min\left(\xi_i  \xi_{\mathcal{N}_i} \hat{A}_i, \text{clip}(\xi_i, 1- \epsilon, 1+\epsilon) \xi_{\mathcal{N}_i} \hat{A}_i \right)  \right.\\
    & - \kappa_i \cdot \sum_{j \in \mathcal{N}_i}  
      \rho_{j}  \mathbb{I}_{X^{ij}}(s, \boldsymbol{a}) \Vert \pi^{ij}(\cdot|s; \theta^{ij}) - \pi^{jj}(\cdot|s)\Vert_2^2 \left. + \rho'_{j} \mathbb{I}_{X^{ii}}(s, \boldsymbol{a})\Vert\pi^{ii}(\cdot|s; \theta^{ii}) - \pi^{ji}(\cdot|s)\Vert_2^2 \right],
\end{aligned}
\end{equation}
where $\theta^{ii}$ denotes the parameters of $\pi^{ii}$ and $ \boldsymbol{\theta}^{-ii}$ denotes the parameters of all the $\pi^{ij}$ ($j \in \mathcal{N}_i$).
With this objective, each agent optimises its own policy $\pi^{ii}$ under the constraint of staying close to the suggested policies. In the meanwhile, the suggestions $\pi^{ij}$ which are involved in $\xi_{\mathcal{N}_i}$, are optimised to maximise the agent's individual advantage function $A_i$ under the constraint of avoiding deviating too far from the actual policies of other agents. 
This objective function balances individual policy updates with the need for coordination among agents, thereby aligning individual objectives with collective goals. 


In our implementation, we use $\hat{\kappa}_i = \text{mean}_{s, \boldsymbol{a}}|\hat{A}_i^{\boldsymbol{\pi}}|$ to approximate $\kappa_i$ in order to mitigate the impact of value overestimation. Additionally, we adopt the same value for the coefficients $\rho_j$ and $\rho'_j$ across different $j$, and denote it as $\rho$. We also utilize the generalized advantage estimator (GAE) \citep{Schulman2016High-dimensionalEstimation} due to its well-known properties to obtain estimates,
\begin{equation}
    \begin{aligned} \label{advantage}
    &\hat{A}_i^t = \sum_{l=0}^\infty (\gamma \lambda)^l \delta_{t+l}^{V_i}, \qquad \delta_{t+l}^{V_i} = r_i^{t+l} + \gamma V_i(s_{t+l+1}) - V_i(s_{t+l}),
\end{aligned}
\end{equation}
where \( V_i \) is approximated by minimising the following loss,
\begin{equation} \label{V_loss}
    \mathcal{L}_{V_i} = \mathbb{E}[(V_i(s_t) - \sum_{l=0}^\infty \gamma^l r_i^{t+l})^2].
\end{equation}
Algorithm \ref{algo} presents the detailed procedure of our practical algorithm. A corresponding illustration figure can be found in Fig.~\ref{illustration} in Appendix \ref{algorithm_appendix}.

\begin{algorithm}[!h] 
\caption{Suggestion-Sharing-based MARL (SS)}
\label{algo}
\begin{algorithmic}
    \STATE {\bf Initialize}: Policy networks $\boldsymbol{\tilde{\pi}}^i = (\pi^{i1}, \cdots, \pi^{iN})$, value networks $V_i, \forall i \in \{1, \cdots, N\}$
    \FOR{episode = 1 to $E$}
        \STATE $\mathcal{D}_i \gets \phi, \forall i$
        \STATE Observe initial state $s_1$
        \STATE {\color{gray}{\# \emph{Interact with the environment}}}
        \FOR{$t=1$ to $T$}
            \STATE Execute action $a_t^i \in \pi^{ii}(\cdot|s_t)$
            \STATE Observe reward $r_t^i$ and next state $s_{t+1}$
            \STATE Store $(s_t, a_t^i, r_t^i, s_{t+1}) \in \mathcal{D}_i$
        \ENDFOR
        \FOR{iteration = 1 to $K$}
            \STATE {\color{gray}{\# \emph{Share with agents}}}
            \FOR{each agent $i$}
                \STATE Share action distributions $[\pi^{ii}_{old}(\cdot|s_1), \cdots, \pi^{ii}_{old}(\cdot|s_T)]$ to neighbors $\{j \in \mathcal{N}_i\}$
                \STATE Share action suggestions $[\pi^{ij}_{old}(\cdot|s_1), \cdots, \pi^{ij}_{old}(\cdot|s_T)]$ to neighbors $\{j \in \mathcal{N}_i\}$
            \ENDFOR
            \FOR{$i = 1$ to $N$}
                \STATE {\color{gray}{\# \emph{Learn policy and value individually}}}
                \STATE Compute advantage estimates $\hat{A}_i^1, \cdots, \hat{A}_i^T$ using Eq~\ref{advantage}
                \STATE Update $\boldsymbol{\tilde{\pi}}^i$ using Eq~\ref{obj}
                \STATE Update $V_i$ using Eq~\ref{V_loss}
                \STATE $\boldsymbol{\tilde{\pi}}^{i}_{old} \gets \boldsymbol{\tilde{\pi}}^i$
                \STATE {\color{gray}{\# \emph{Update sharing with agents}}}
                \STATE Share action distributions $[\pi^{ii}_{old}(\cdot|s_1), \cdots, \pi^{ii}_{old}(\cdot|s_T)]$ to neighbors $\{j \in \mathcal{N}_i\}$
                \STATE Share action suggestions $[\pi^{ij}_{old}(\cdot|s_1), \cdots, \pi^{ij}_{old}(\cdot|s_T)]$ to neighbors $\{j \in \mathcal{N}_i\}$
            \ENDFOR
        \ENDFOR
    \ENDFOR
\end{algorithmic}
\end{algorithm}


\section{Experimental Settings and Results} \label{sec:results}

We evaluate our method with four diverse environments where agents have conflicting individual rewards. Three environments are adapted from related works, while we propose one of our own environment to facilitate the analysis of the problem and the performance of our method.

\subsection{Environments}

We evaluate our approach in diverse environments designed to capture distinct cooperation and dilemma scenarios. The environments are described below:

\textbf{Cleanup}. This environment represents a public goods dilemma, adopted from the setting in \citep{Christoffersen2023GetRL}. Agents must clean a river and eating apples. Apples spawn only if the river's waste density is below a threshold, with the spawn rate inversely proportional to the waste density. Eating an apple rewards an agent with $+1$, while cleaning the river provides neither a reward nor a cost. This setup creates a free-rider problem, where agents may prioritise eating apples over cleaning the river, potentially undermining collective performance. For efficiency, we reduce the environment size to $11\times 18$ and the episode time horizon to $100$ time steps, smaller than in \citep{Christoffersen2023GetRL}, to decrease training time. 

\textbf{Harvest}. This environment represents a tragedy of the commons dilemma, where agents harvest apples in a shared space. Based on \citep{Christoffersen2023GetRL}, apples spawn at a rate proportional to the number of apples around the spawn positions. Only eating an apple provides a reward of $+1$. The challenge is for agents to harvest apples sustainably while collaborating to avoid over-harvesting in the same region. To reduce training time, we set the  episode time horizon to $100$ time steps and environment size to $7 \times 38$, both smaller than in \citep{Christoffersen2023GetRL}.

\textbf{Cooperative navigation (C. Navigation)}.  In this environment, each agent must navigate to a designated landmark. We use the same observation and action configurations as in \citep{Zhang2018FullyAgents}. Agents earn rewards based on their proximity to targets but incur a $-1$ penalty for collisions. Communication is limited to adjacent agents. We set the time horizon of an episode as 100 time steps and use three agents. The environment size is $5 \times 5$, with three agents and an episode time horizon of 100 time steps. Fig.~\ref{environments}(a) in Appendix~\ref{appendix_environment} illustrates the setup. 

\textbf{Cooperative predation (C. Predation)}. This environment involves a sequential social dilemma in a continuous domain, where multiple predator agents aim to capture a single prey. All predators cooperating (approaching the prey) results in each receiving a reward of $-1$. Universal defection (not approaching) yields a reward of $-2N+1$ for each predator, where $N$ is the total number of agents. In mixed scenarios, predators pursuing the prey receive a reward of $-2N$, while non-participating predators gain $0$. The challenge is to incentivise agents to cooperate and capture the prey rather than acting selfishly. At the start of each episode, the prey's position, $x_{tar} \in \mathcal{X}$, and the agents' initial positions, $x_{ag_i} \in \mathcal{X}$, are randomly assigned within $\mathcal{X} = [0, 30]$. The state is represented as $s^t = [x_{ag_1}^t - x_{tar}, \ldots, x_{ag_N}^t - x_{tar}]$, a continuous variable. The action set $\mathcal{A} = \{-1, +1\}$ corresponds to left and right movements. Neighbouring agents are defined as those within a normalised distance of $0.1$. Fig.~\ref{environments}(b) in Appendix \ref{appendix_environment} illustrates this environment. The episode time horizon is set to $30$, and our main experiments use $8$ predator agents.

\subsection{Baselines}

We evaluate our SS framework against five baseline algorithms designed to optimise the collective return of all agents under individual rewards, ensuring a fair comparison that highlights SS's competitiveness without relying on value or policy sharing. While many other MARL algorithms are commonly used as baselines in the literature, we exclude them due to fundamental differences in problem settings.

To ensure comparability, all baseline algorithms and our SS algorithm are built on the same PPO-based MARL framework. This ensures that observed performance differences arise from the information-sharing mechanisms rather than underlying algorithmic variations. The hyperparameters used in the experiments are detailed in Appendix \ref{hyperparameters} and were selected based on standard practices in the field. For example, we set the discount factor to $0.99$ and used the same clipping threshold as in the original PPO paper \citep{Schulman2015TrustOptimization}. Network sizes were tailored to the state and action dimensions of each environment. 

\textbf{Value Function Parameter Sharing (VPS)} \citep{Zhang2018FullyAgents}: This approach employs a consensus method to update individual value functions. Each update utilises the agent's unique reward while incorporating a weighted aggregation of value function parameters from neighbouring agents.

\textbf{Value Sharing (VS)} \citep{Du2022ScalableSystems}: In this method, each agent independently learns a value function and shares the output values with its neighbours. The individual policy network is then updated based on the average of the shared values.

\textbf{Policy Parameter Sharing (PS)} \citep{Zhang2019DistributedConsensus}: This algorithm uses consensus updates to learn policies for all agents. Each agent learns $N$ policies based on individual rewards and aggregates policy parameters with neighbours. Value functions, however, are learned independently without consensus updates.

\textbf{Centralized Learning (CL)}: In this method, a centralised value function is learned based on the sum of individual rewards, while each agent learns an individual policy. To avoid the high dimensionality of joint action spaces, a single policy for joint actions is not employed.

\textbf{Intrinsic Moral Rewards (IMR)}: This approach provides intrinsic rewards to cooperative agents in addition to environmental rewards, based on the virtue-kindness moral type proposed in \citep{Tennant2023ModelingLearning}. Each agent learns independently using both individual external rewards and IMR. However, performance is evaluated based solely on external rewards to ensure comparability with other algorithms. Specifically, in Cleanup, IMR rewards an agent for cleaning the river. In Harvest, an agent receives IMR for abstaining from eating apples. In C. Predation, IMR is given to each agent that approaches the prey. For C. Navigation, applying IMR is challenging because cooperative behaviour is not tied to specific actions.

It is important to note that CL requires a centralised learning unit, and IMR involves additional rewards, which may limit their practical feasibility. Nonetheless, we include these methods in the baselines to provide a comprehensive comparison for evaluating the performance of our algorithm.

\subsection{Experimental Results}

\begin{figure}[t]
\centering
\begin{subfigure}[t]{0.244\linewidth}
    \centering
    \includegraphics[width=\linewidth]{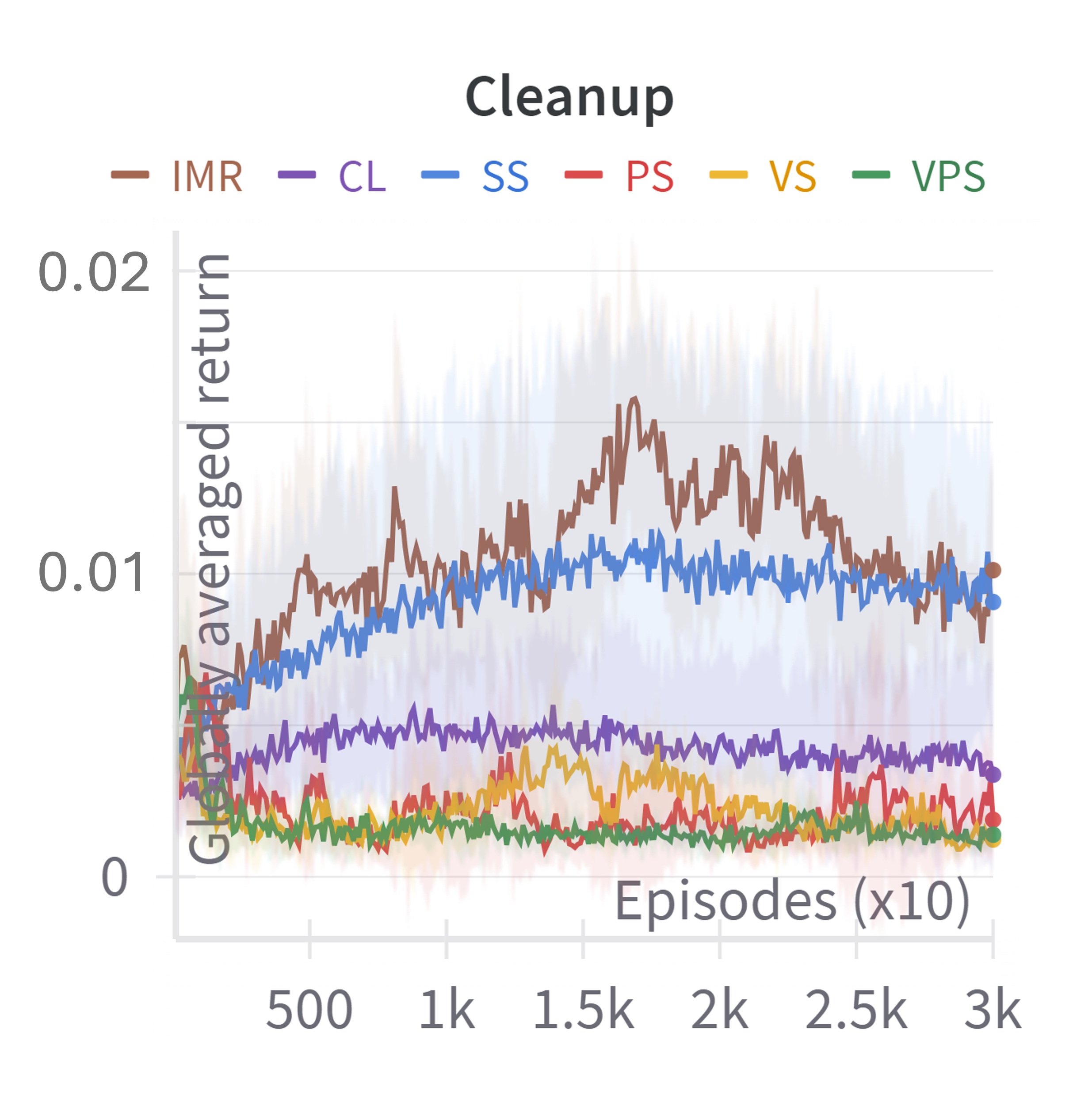}
    \caption{Cleanup}
\end{subfigure}
\hfill
\begin{subfigure}[t]{0.244\linewidth}
    \centering
    \includegraphics[width=\linewidth]{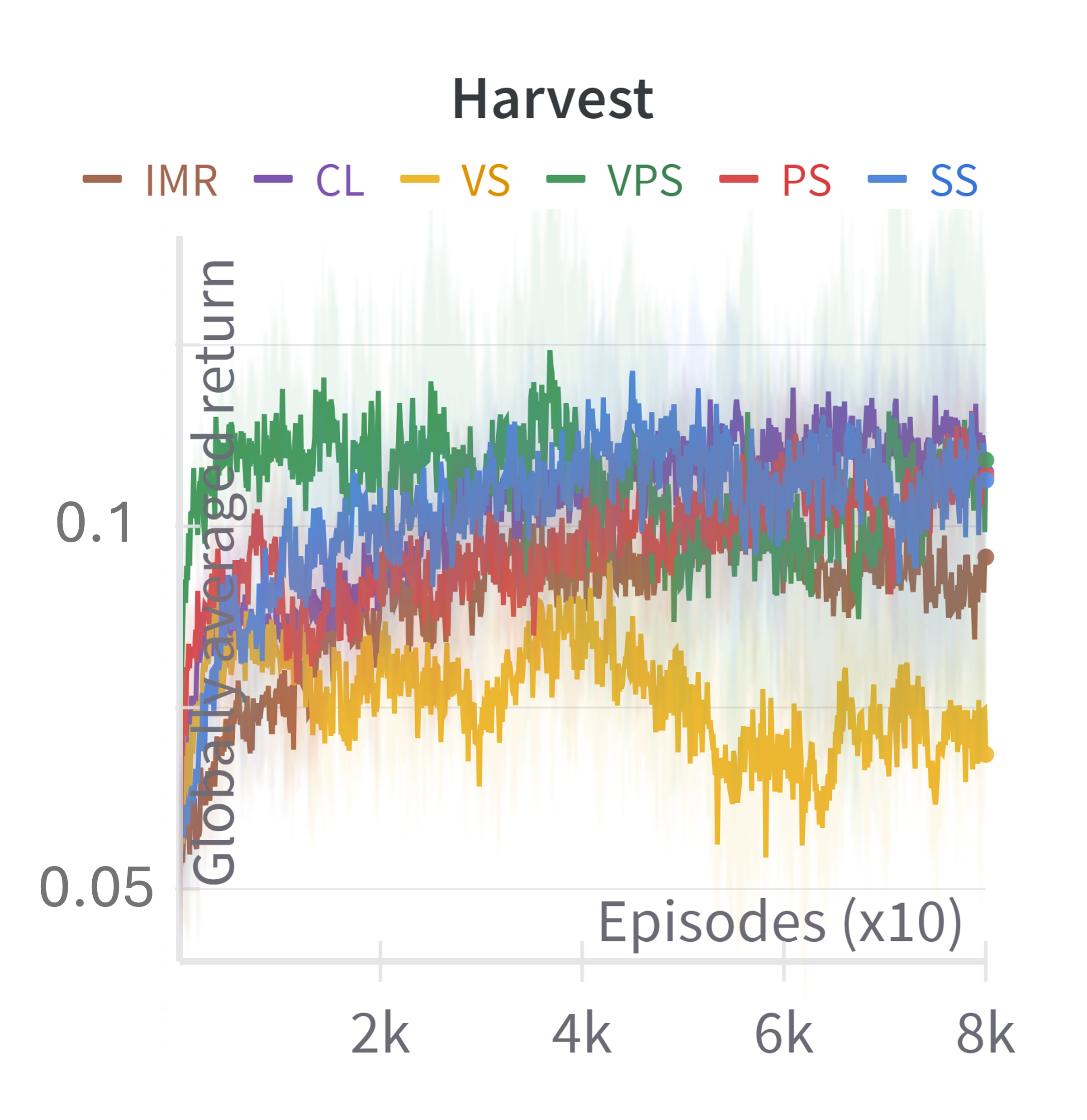}
    \caption{Harvest}
\end{subfigure}
\hfill
\begin{subfigure}[t]{0.244\linewidth}
    \centering
    \includegraphics[width=\linewidth]{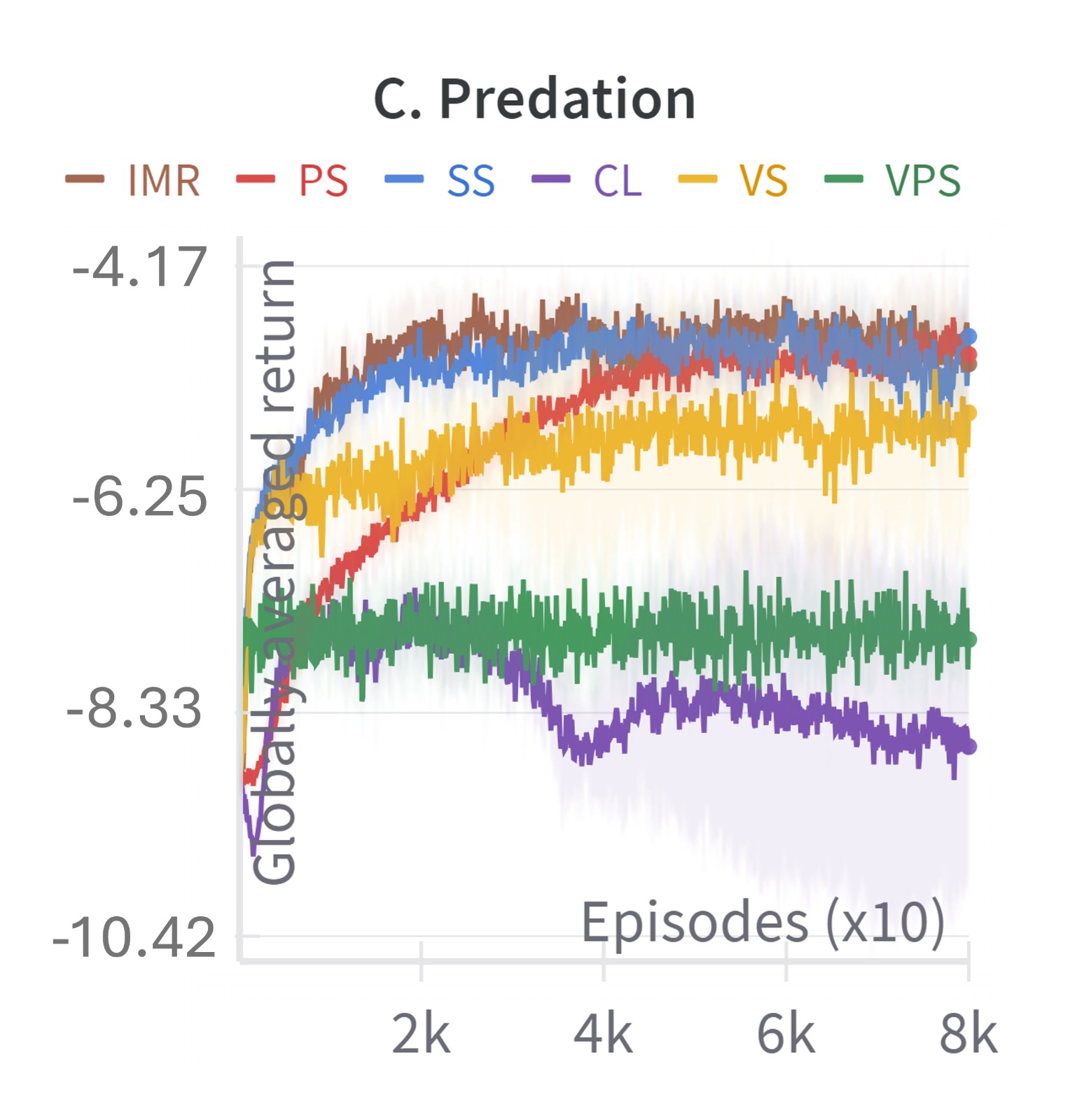}
    \caption{C.Predation}
\end{subfigure}
\hfill
\begin{subfigure}[t]{0.244\linewidth}
    \centering
    \includegraphics[width=\linewidth]{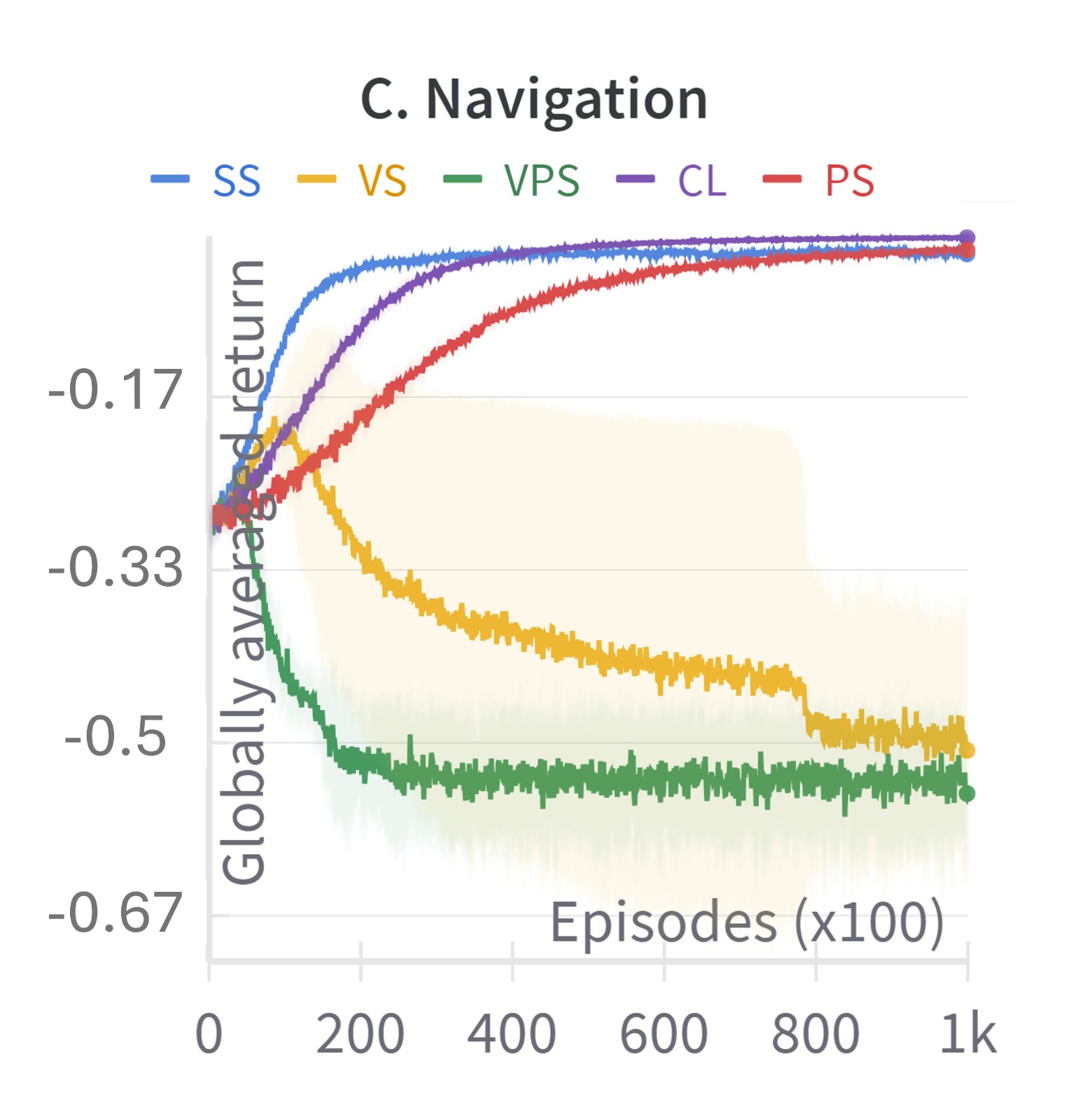}
    \caption{C.Navigation}
\end{subfigure}
\caption{Training curves of globally averaged return.}
\label{main_results_curves}
\end{figure}

\begin{figure}[t]
\centering
\begin{subfigure}[t]{0.235\linewidth}
    \centering
    \includegraphics[width=\linewidth]{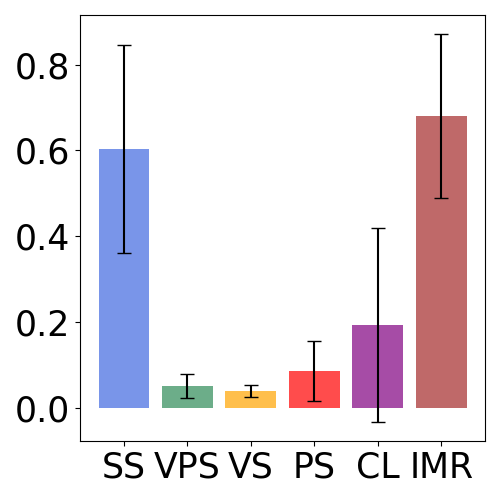}
    \caption{Cleanup}
\end{subfigure}
\hfill
\begin{subfigure}[t]{0.235\linewidth}
    \centering
    \includegraphics[width=\linewidth]{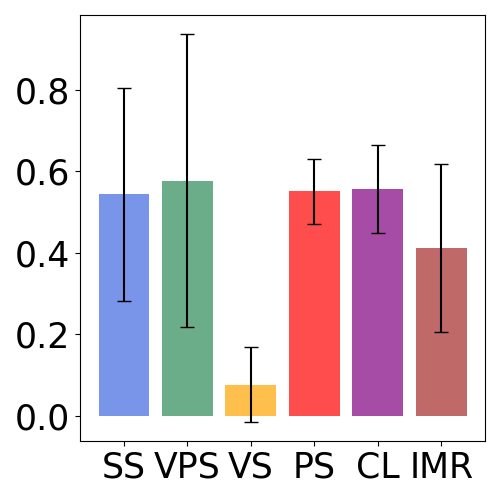}
    \caption{Harvest}
\end{subfigure}
\hfill
\begin{subfigure}[t]{0.235\linewidth}
    \centering
    \includegraphics[width=\linewidth]{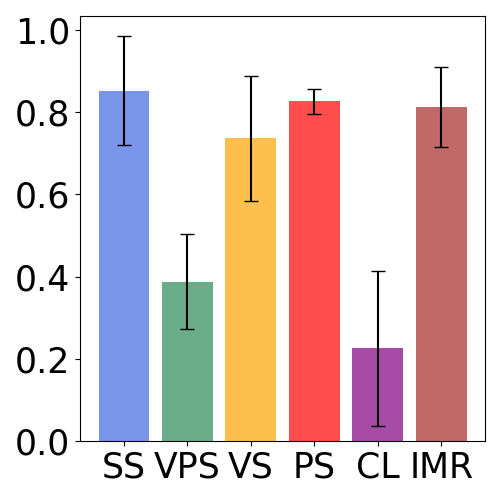}
    \caption{C.Predation}
\end{subfigure}
\hfill
\begin{subfigure}[t]{0.235\linewidth}
    \centering
    \includegraphics[width=\linewidth]{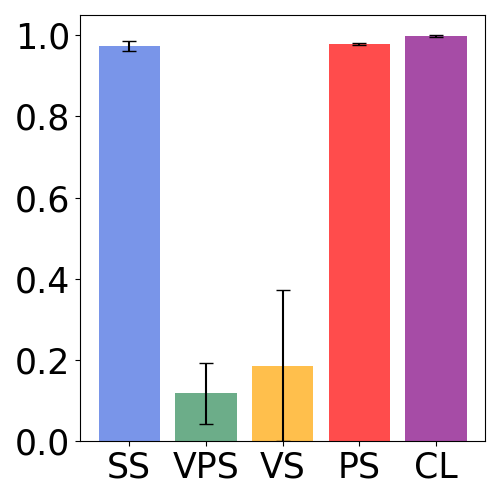}
    \caption{C.Navigation}
\end{subfigure}
\caption{Final results of normalised globally averaged return.}
\label{main_results_bars}
\end{figure}

\textbf{Main results.} We conducted 5 runs with different seeds for each algorithm and environment. Fig.~\ref{main_results_curves} shows the training curves and Fig.~ \ref{main_results_bars} the normalised final averaged returns for different algorithms. The averaged return refers to the collective return, normalised by the number of agents and  episode length. Our SS algorithm demonstrates consistently strong performance across all tasks, with averaged returns matching or exceeding those of baseline algorithms that rely on sharing values or policy parameters. This shows that SS is an effective method of learning cooperative policies for collective return by sharing suggestions instead of values or policies.

In Fig.~\ref{main_results_curves}, SS converges faster than PS, which implies that sharing action distributions is more efficient than sharing parameters of policy networks. In Fig.~\ref{main_results_bars}, SS outperforms both VS and VPS in almost all the tasks. Additionally, PS shows better performance than VS and VPS, which may indicate that sharing policy information is more effective than sharing value information. Notably, SS outperforms CL in some cases. We hypothesise that in these scenarios, SS facilitates cooperation by enabling agents to encourage each other through action suggestions based on their individual interests, while CL may struggle due to exploration issues arising from a lack of successful cooperation experiences. IMR also shows competitive performance, even achieving the best results in a specific case. However, for the problem addressed in this paper, adding intrinsic rewards may not always be practical, especially in scenarios where designing appropriate intrinsic rewards is challenging.


\begin{figure}[t]
\centering
\begin{subfigure}[t]{0.244\linewidth}
    \centering
    \includegraphics[width=\linewidth]{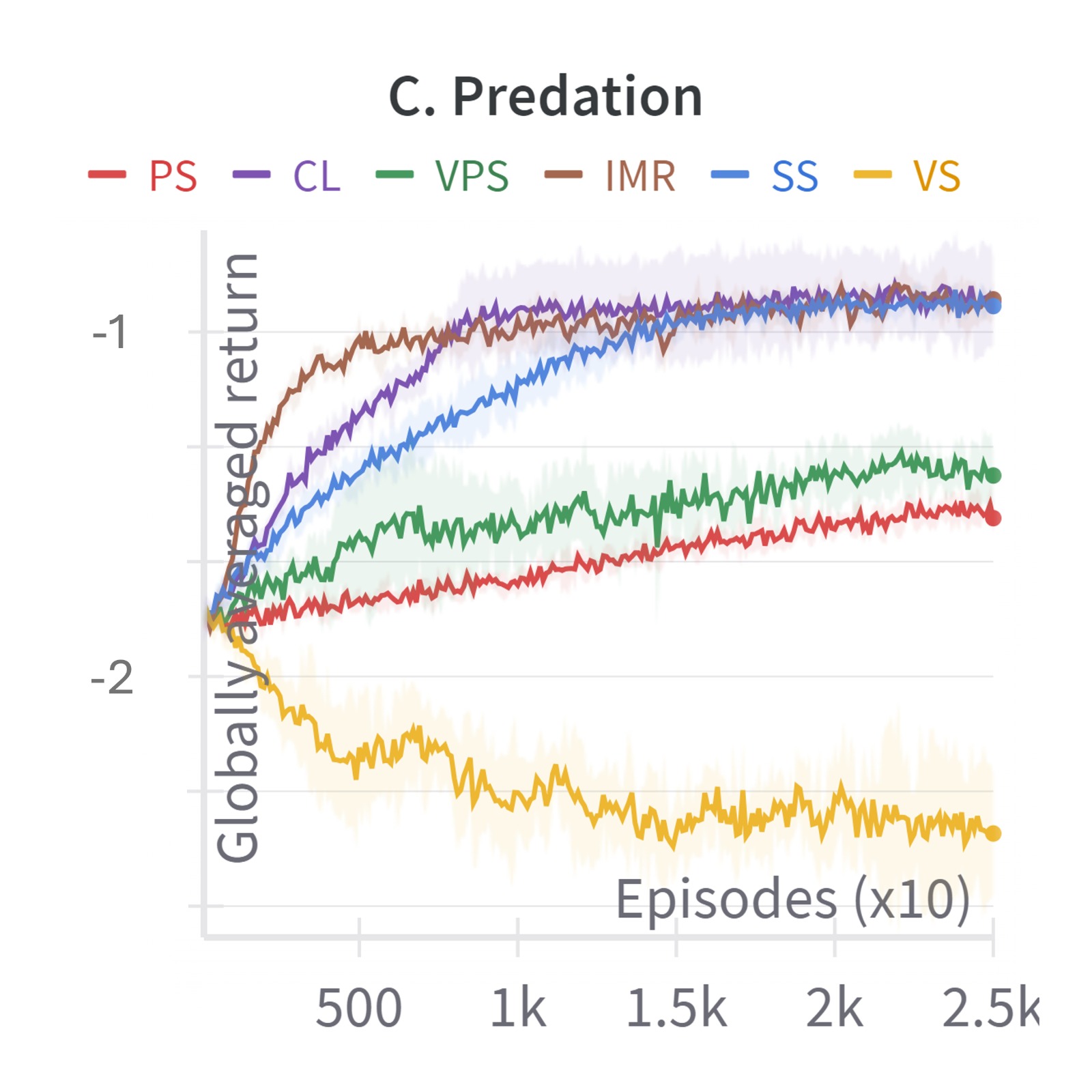}
    \caption{Averaged return}
\end{subfigure}
\hfill
\begin{subfigure}[t]{0.244\linewidth}
    \centering
    \includegraphics[width=\linewidth]{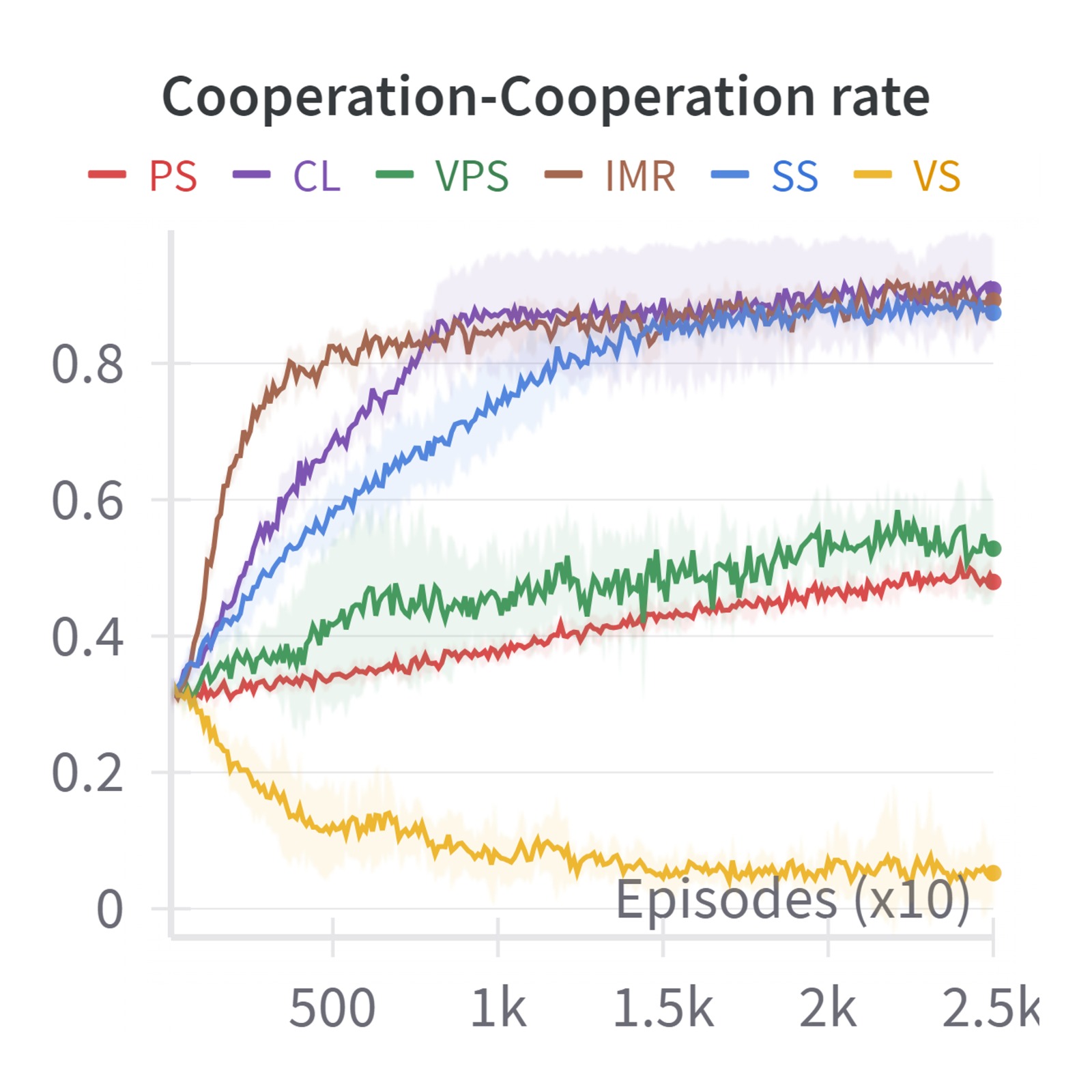}
    \caption{C-C rate}
\end{subfigure}
\hfill
\begin{subfigure}[t]{0.244\linewidth}
    \centering
    \includegraphics[width=\linewidth]{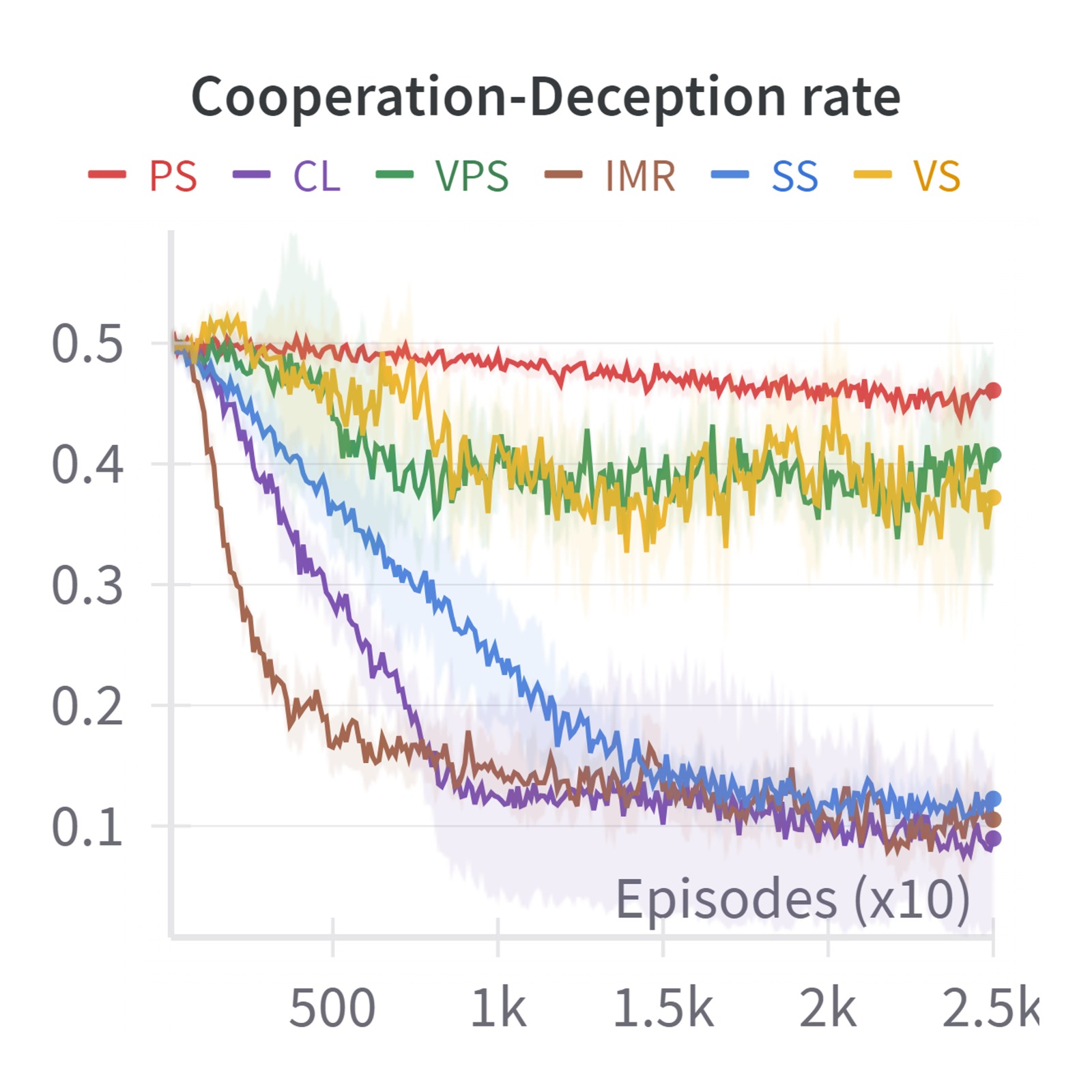}
    \caption{C-D rate}
\end{subfigure}
\hfill
\begin{subfigure}[t]{0.244\linewidth}
    \centering
    \includegraphics[width=\linewidth]{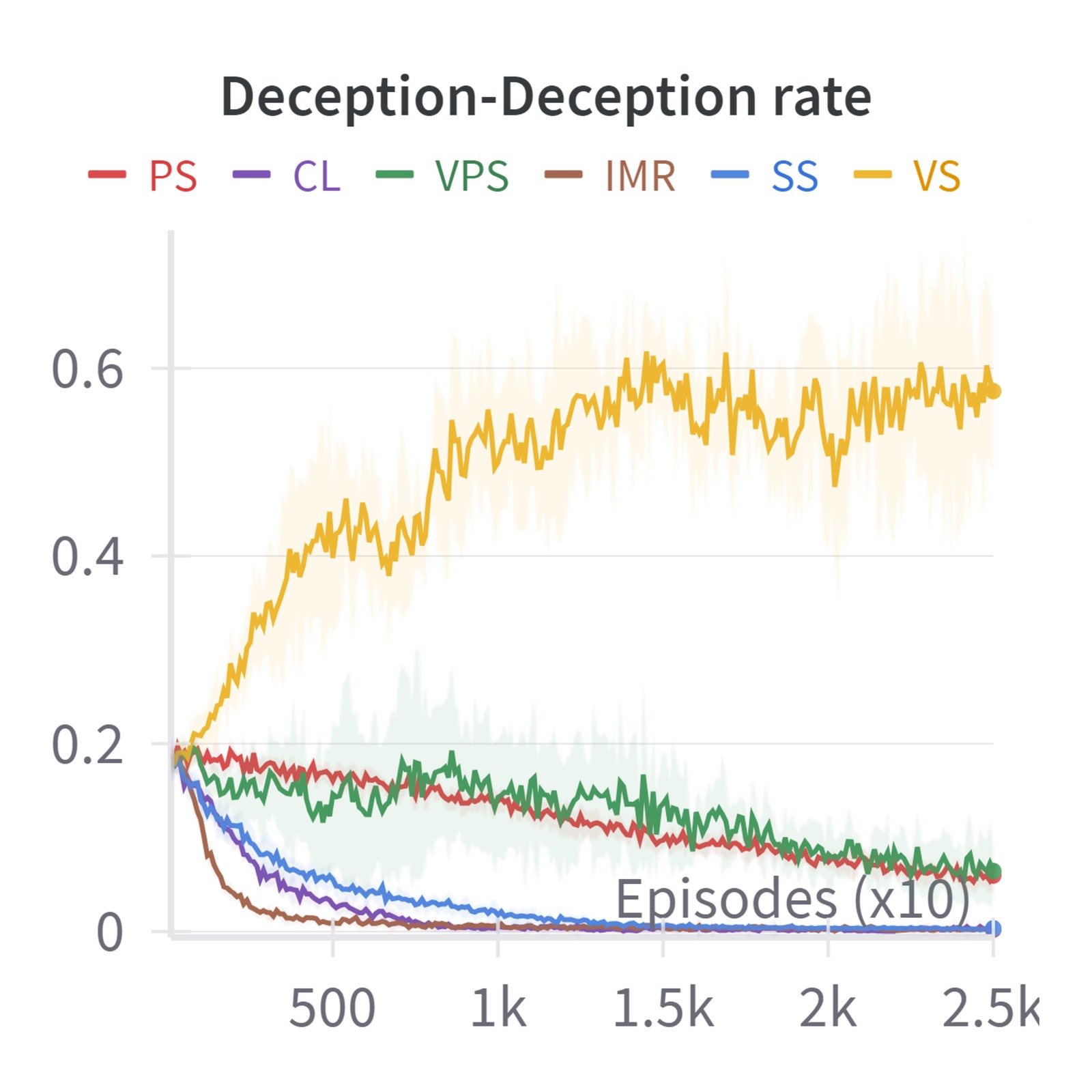}
    \caption{D-D rate}
\end{subfigure}
\caption{Analytical results on C. Predation with two agents.}
\label{fig:analytical_res}
\end{figure}

\textbf{Effect on solving sequential social dilemmas.} SS is designed to address scenarios where agents' conflicting individual interests hinder collective cooperation, such as in Social Dilemmas. The C. Predation task, an extension of the sequential Prisoner's Dilemma, clearly illustrates the effectiveness of SS in managing these conflicting interests. In the C. Predation task, the selfish policy for each agent is to defect (act as a free rider) by not moving towards the prey. However, the collectively optimal solution requires all agents to cooperate by moving towards the prey. Fig.~\ref{fig:analytical_res} shows results for two agents, with sub-figures (b)-(d) presenting statistical data on the rates of each type of joint action: both agents cooperating and moving towards the prey (C-C), one agent moving towards the prey while the other defects (C-D), and both agents defecting by moving away from the prey (D-D). As shown in the results, SS converges to optimal cooperative policies, achieving a C-C rate close to 1. This highlights SS's ability to foster cooperation, overcoming the challenges posed by the Prisoner's Dilemma in a sequential setting. It effectively aligns agents' actions towards the collective goal, despite individual incentives to defect.

\begin{figure}[t]
    \centering
    \begin{subfigure}[t]{0.3\linewidth}
        \centering
        \includegraphics[width=\linewidth]{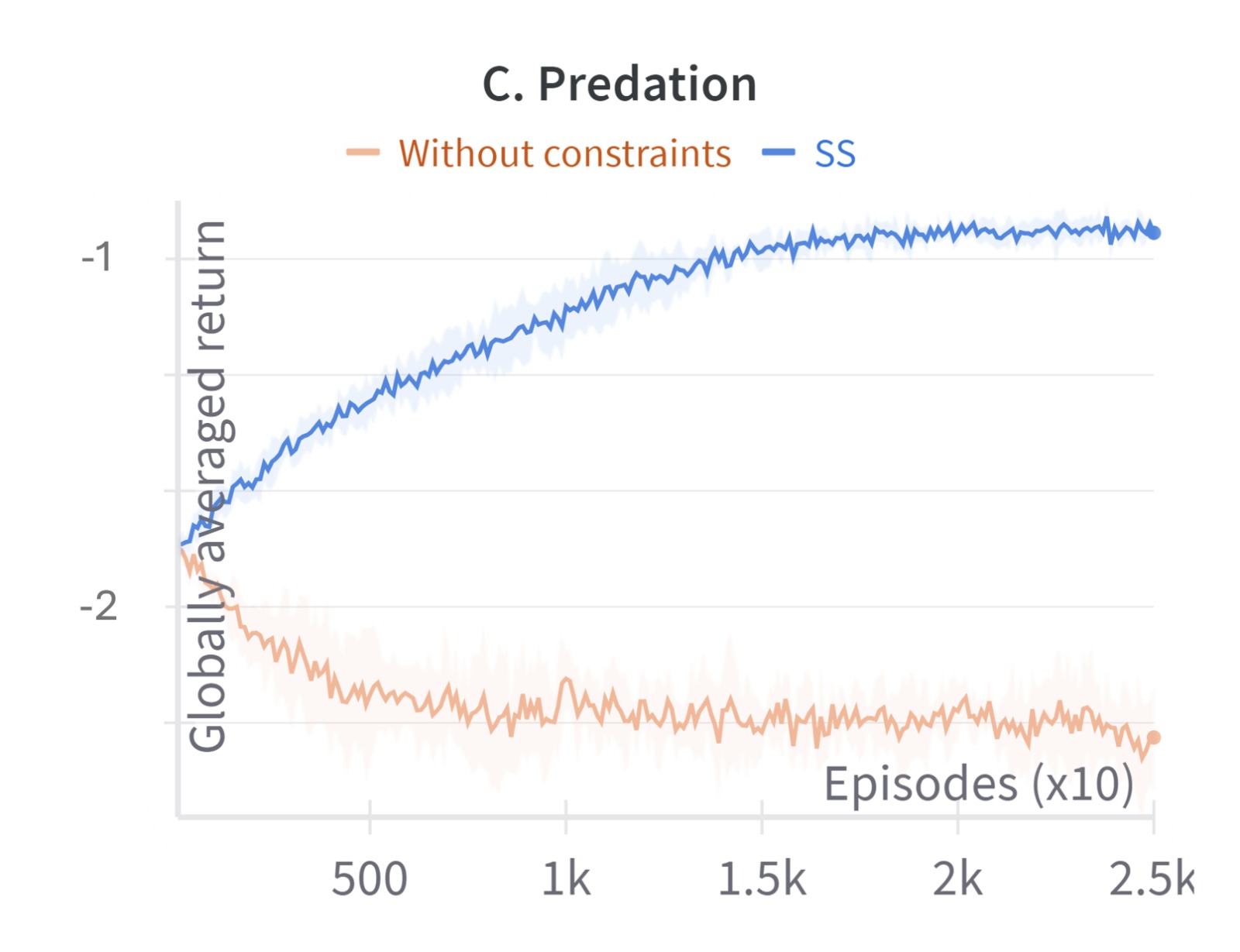}
        \caption{C. Predation}
    \end{subfigure}
    \begin{subfigure}[t]{0.3\linewidth}
        \centering
        \includegraphics[width=\linewidth]{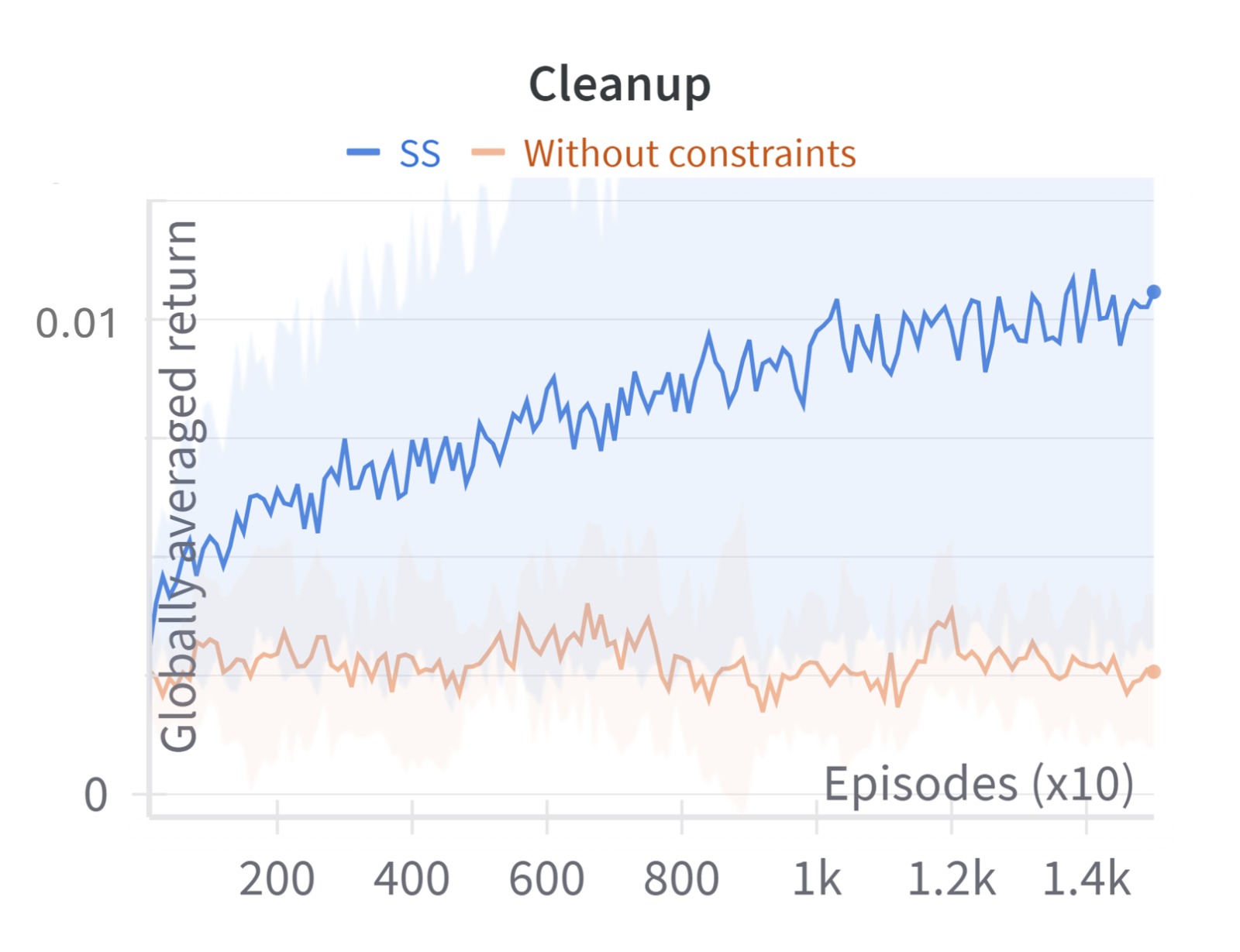}
        \caption{Cleanup}
    \end{subfigure}
    \caption{Ablation study of removing constraints.}
    \label{fig_ablation}
\end{figure}

\textbf{Ablation study involving objective constraints.} We conducted an ablation study by removing the constraints in the objective function, i.e., setting $\rho = 0$. The experimental results, shown in Fig.~\ref{fig_ablation}, indicate that removing the constraints leads to a significant drop in algorithm performance. This demonstrates that shared policy suggestions are essential for learning optimal collective policies. Without incorporating these shared suggestions to guide individual policy learning, agents fail to learn how to maximise collective returns.

\begin{figure}[t]
    \centering
    \begin{subfigure}[t]{0.244\linewidth}
        \centering
        \includegraphics[width=\linewidth]{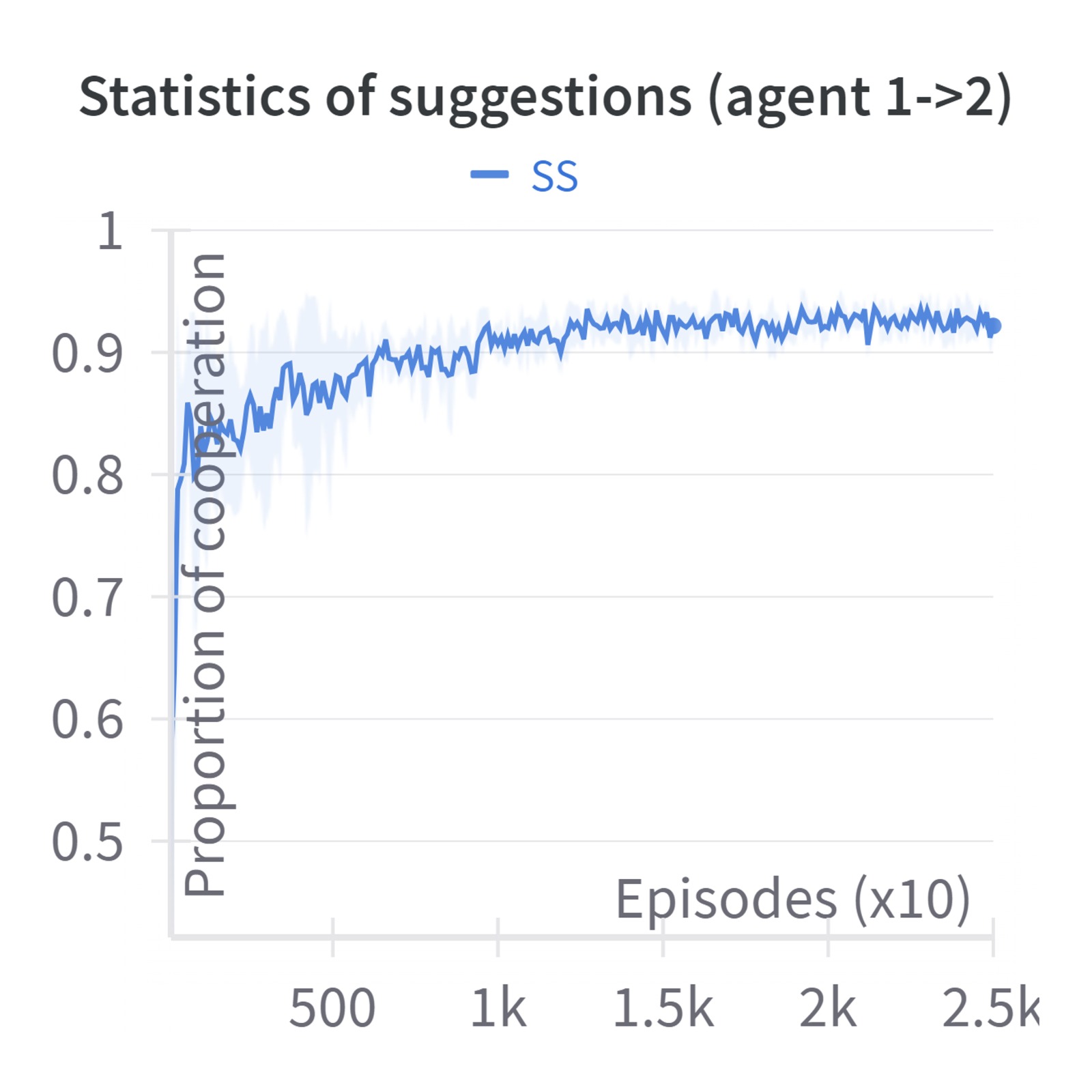}
        \caption{}
    \end{subfigure}
    \begin{subfigure}[t]{0.244\linewidth}
        \centering
        \includegraphics[width=\linewidth]{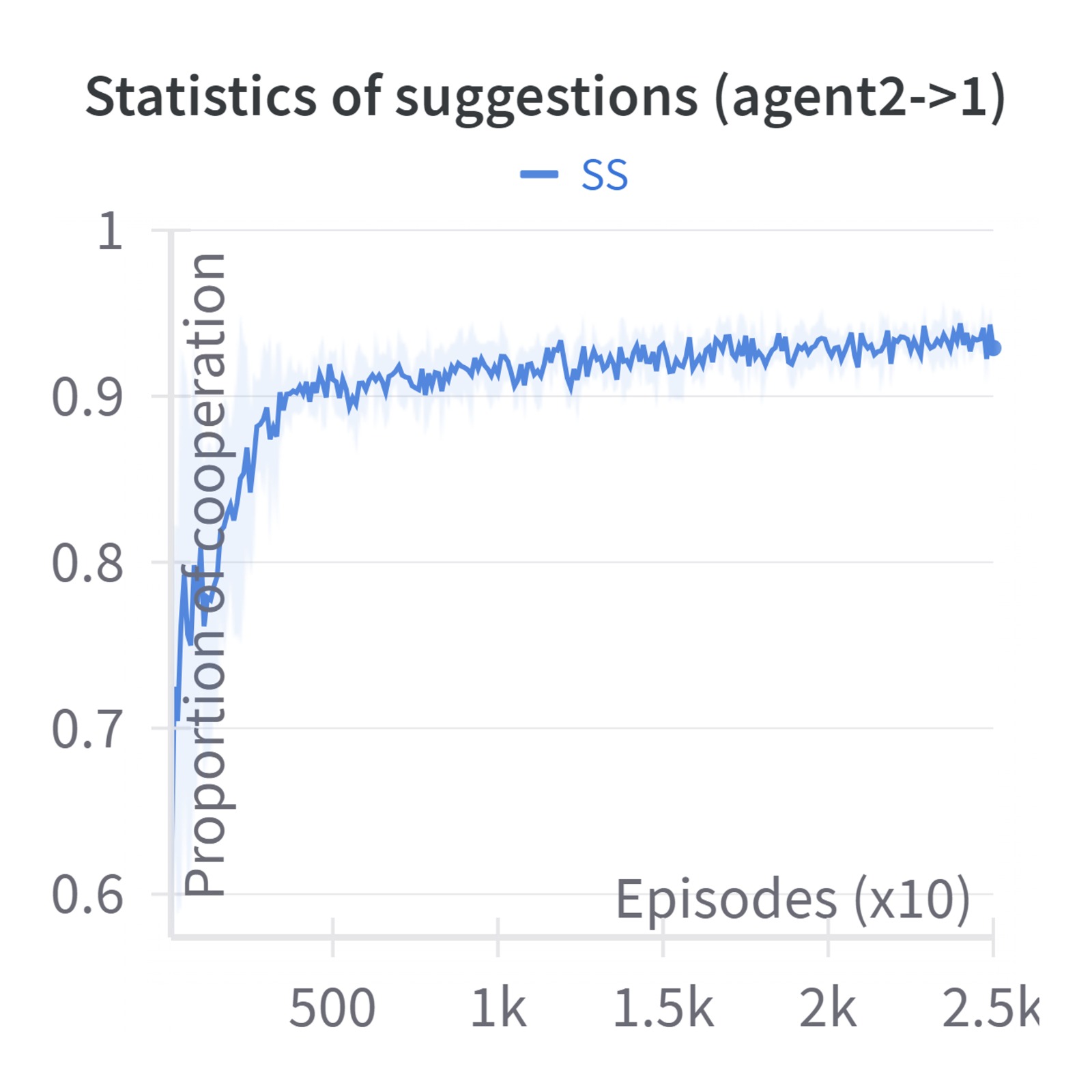}
        \caption{}
    \end{subfigure}
    \begin{subfigure}[t]{0.244\linewidth}
        \centering
        \includegraphics[width=\linewidth]{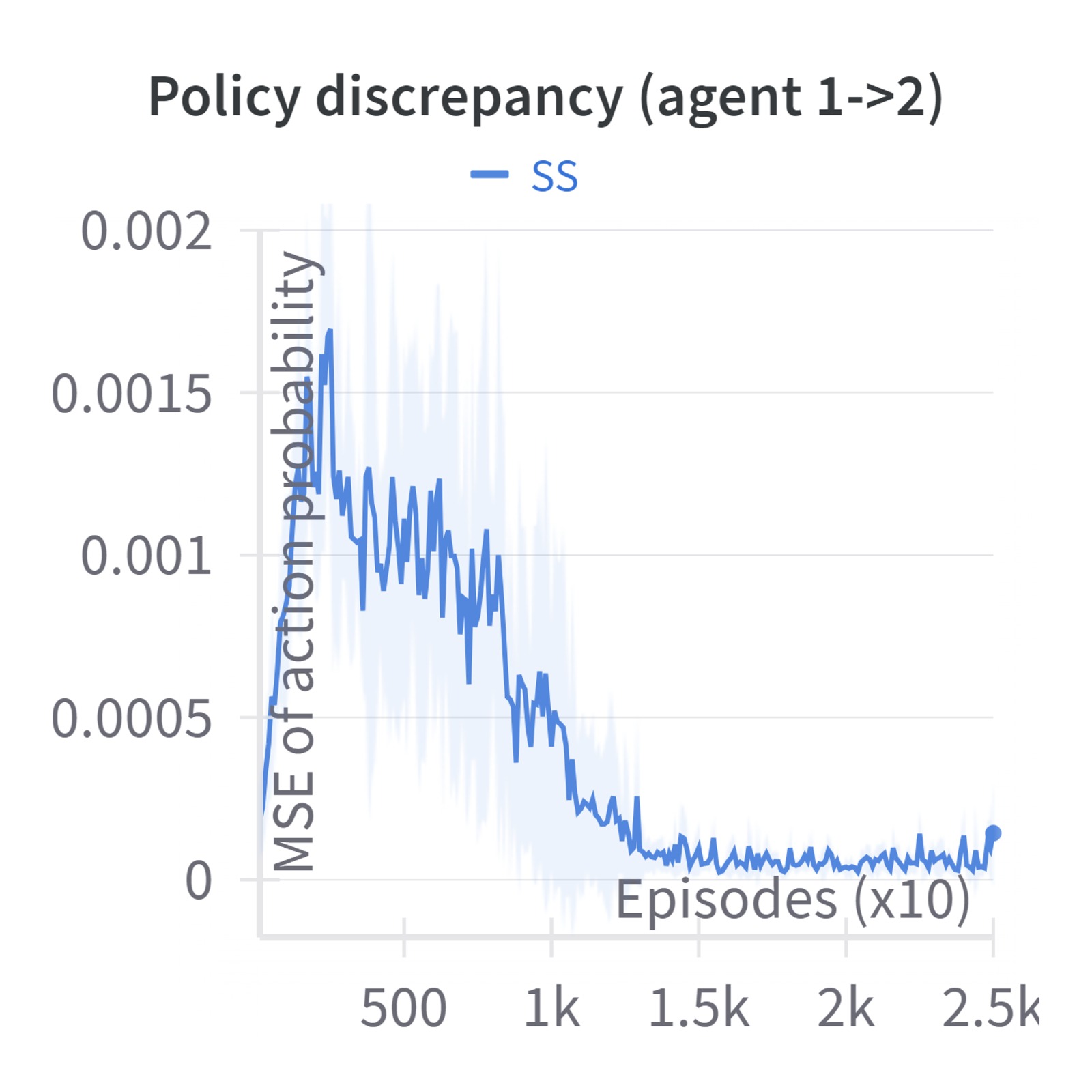}
        \caption{}
    \end{subfigure}
    \begin{subfigure}[t]{0.244\linewidth}
        \centering
        \includegraphics[width=\linewidth]{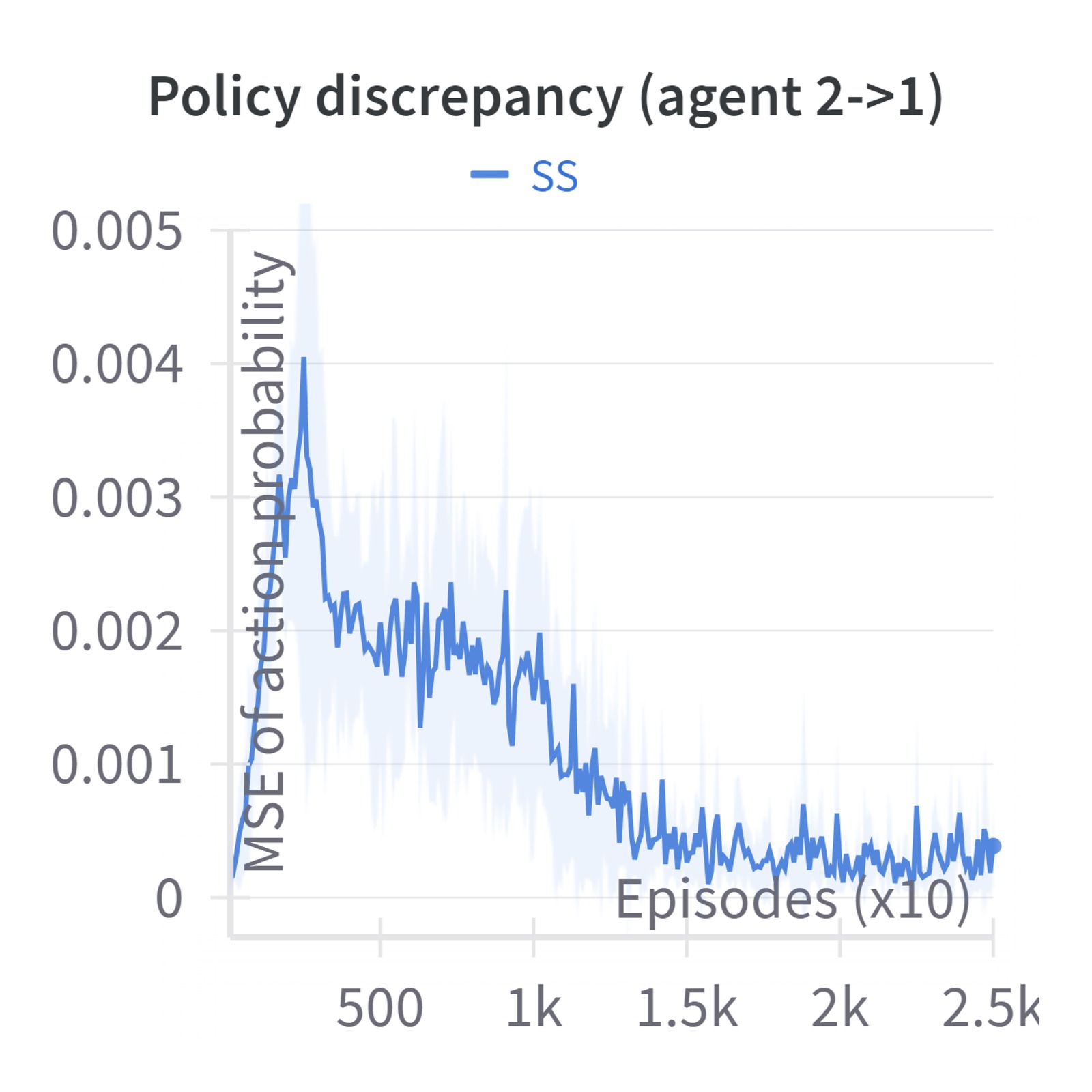}
        \caption{}
    \end{subfigure}
    \caption{Statistics of suggestions and discrepancy.}
    \label{fig_metrics}
\end{figure}

\textbf{Policy suggestion and policy discrepancy.}
We conducted experiments to investigate the learned policy suggestions and the discrepancy between an agent's policy and the suggested policy given by another agent. For clarity, we used the task of C. Predation with two agents. In this task, the action set included two actions: ``moving towards the target" and ``moving away from the target." The optimal policy to maximise the collective total returns was for both agents to move towards the target.
To examine the policy suggestions learned by each agent, we calculated the proportion of suggested actions that were ``moving towards the target." The results, shown in Fig.~\ref{fig_metrics} (a) and (b), indicated that both agents learned to suggest the other agent move towards the target with a proportion approaching 1. The mean square error (MSE) between the probability of the action chosen by an agent and the suggested action given by the other agent is shown in Fig.~\ref{fig_metrics} (c) and (d). As training progressed, the MSE decreased and approached 0.

\section{Discussion and Conclusion} \label{sec:conclusions}

In this work, we addressed the challenge of achieving collective welfare in scenarios where individual interests may conflict with collective objectives. We proposed a Suggestion-Sharing-based MARL method, designed for situations where agents lack access to others' rewards and policies, and traditional methods relying on sharing rewards, values, or policy models are infeasible. SS enables agents to incorporate their individual interests into action suggestions for other agents. Taking into account the suggestions shared by others when learning individual policies can facilitate implicit inferences about collective interests and then facilitate learning policies that can promote collective welfare.

Theoretically, we demonstrated that the discrepancy between agents' action distributions and the suggestions they receive bounds the difference between individual and collective objectives. This theoretical insight led to a novel optimisation problem, decomposable into individual agents' objectives, which serves as a lower bound for the original collective goal. Iteratively solving these decomposed problems drives agents toward cooperative behaviours. Empirically, our experiments showed that SS achieves competitive performance compared to baseline algorithms that rely on sharing value functions, policy parameters, or intrinsic rewards.

Despite its promising results, SS has several limitations and opens up directions for future work. First, the current implementation of SS requires training $N^2$ policy networks, as each agent learns its own policy and suggests policies for other agents. This raises scalability challenges for larger systems. Future work could address this by employing more computationally efficient architectures, such as multi-head policy networks with $N$ outputs: one for the agent's own policy and $N-1$ for the suggested policies for others. Second, SS assumes that agents truthfully share their suggestions. However, in practical scenarios, agents may act selfishly or deceptively. This limitation motivates future research on incorporating mechanisms to handle varying levels of trust, such as reputation systems or incentive structures to encourage truthful sharing of suggestions. Third, while SS avoids explicit sharing of rewards, values, or policies, it does not provide formal privacy guarantees. This work qualitatively reduces information sharing compared to methods that directly share rewards or full policies, but it does not minimise information leakage quantitatively. Future research could explore techniques to enhance privacy guarantees while maintaining cooperative performance, such as leveraging cryptographic approaches or differential privacy.

In summary, SS represents an important step toward achieving multi-agent cooperation for collective welfare, offering a performant and privacy-conscious approach to MARL. By addressing its current limitations, SS has the potential to further advance the field of cooperative multi-agent systems.

\subsection*{Acknowledgments}
Giovanni Montana acknowledges support from a UKRI AI Turing Acceleration Fellowship (EPSRC EP/V024868/1).

\bibliography{main}

\begin{thebibliography}{53}
\providecommand{\natexlab}[1]{#1}
\providecommand{\url}[1]{\texttt{#1}}
\expandafter\ifx\csname urlstyle\endcsname\relax
  \providecommand{\doi}[1]{doi: #1}\else
  \providecommand{\doi}{doi: \begingroup \urlstyle{rm}\Url}\fi

\bibitem[Albrecht and Stone(2018)]{Albrecht2018AutonomousProblems}
Stefano~V. Albrecht and Peter Stone.
\newblock {Autonomous agents modelling other agents: A comprehensive survey and open problems}.
\newblock \emph{Artificial Intelligence}, 258\penalty0 (September):\penalty0 66--95, 2018.
\newblock ISSN 00043702.
\newblock \doi{10.1016/j.artint.2018.01.002}.

\bibitem[Chen et~al.(2022)Chen, Zhang, Giannakis, and Basar]{Chen2022Communication-EfficientLearning}
Tianyi Chen, Kaiqing Zhang, Georgios~B. Giannakis, and Tamer Basar.
\newblock {Communication-Efficient Policy Gradient Methods for Distributed Reinforcement Learning}.
\newblock \emph{IEEE Transactions on Control of Network Systems}, 9\penalty0 (2):\penalty0 917--929, 2022.
\newblock ISSN 23255870.
\newblock \doi{10.1109/TCNS.2021.3078100}.

\bibitem[Christoffersen et~al.(2023)Christoffersen, Haupt, and Hadfield-Menell]{Christoffersen2023GetRL}
Phillip J.~K. Christoffersen, Andreas~A. Haupt, and Dylan Hadfield-Menell.
\newblock {Get It in Writing: Formal Contracts Mitigate Social Dilemmas in Multi-Agent RL}.
\newblock \emph{Proceedings of the 2023 International Conference on Autonomous Agents and Multiagent Systems}, pages 448--456, 2023.
\newblock URL \url{http://arxiv.org/abs/2208.10469}.

\bibitem[Chu et~al.(2020{\natexlab{a}})Chu, Chinchali, and Katti]{Chu2020Multi-agentControlb}
Tianshu Chu, Sandeep Chinchali, and Sachin Katti.
\newblock {Multi-agent Reinforcement Learning for Networked System Control}.
\newblock \emph{International Conference on Learning Representations}, \penalty0 (1), 2020{\natexlab{a}}.
\newblock URL \url{http://arxiv.org/abs/2004.01339}.

\bibitem[Chu et~al.(2020{\natexlab{b}})Chu, Wang, Codec{\`{a}}, and Li]{Chu2020Multi-AgentControl}
Tianshu Chu, Jie Wang, Lara Codec{\`{a}}, and Zhaojian Li.
\newblock {Multi-Agent Deep Reinforcement Learning for Large-Scale Traffic Signal Control}.
\newblock \emph{IEEE Transactions on Intelligent Transportation Systems}, 21\penalty0 (3):\penalty0 1086--1095, 2020{\natexlab{b}}.
\newblock ISSN 15582914.

\bibitem[Du et~al.(2022)Du, Ma, Liu, Lin, Dong, Wang, and Yang]{Du2022ScalableSystems}
Yali Du, Chengdong Ma, Yuchen Liu, Runji Lin, Hao Dong, Jun Wang, and Yaodong Yang.
\newblock {Scalable Model-based Policy Optimization for Decentralized Networked Systems}.
\newblock \emph{International Conference on Intelligent Robots and Systems (IROS)}, pages 9019--9026, 2022.
\newblock URL \url{http://arxiv.org/abs/2207.06559}.

\bibitem[Foerster et~al.(2017)Foerster, Nardell, Farquhar, Afouras, Torr, Kohli, and Whiteson]{Foerster2017StabilisingLearning}
Jakob Foerster, Nantas Nardell, Gregory Farquhar, Trtantafyllos Afouras, Philip~H.S. Torr, Pushmeet Kohli, and Shimon Whiteson.
\newblock {Stabilising experience replay for deep multi-agent reinforcement learning}.
\newblock \emph{34th International Conference on Machine Learning, ICML 2017}, 3:\penalty0 1879--1888, 2017.

\bibitem[Gersani et~al.(2001)Gersani, Brown, O'Brien, Maina, and Abramsky]{Gersani2001TragedyCompetition}
Mordechai Gersani, Joel~S. Brown, Erin~E. O'Brien, Godfrey~M. Maina, and Zvika Abramsky.
\newblock {Tragedy of the commons as a result of root competition}.
\newblock \emph{Journal of Ecology}, 89\penalty0 (4):\penalty0 660--669, 2001.
\newblock ISSN 00220477.
\newblock \doi{10.1046/j.0022-0477.2001.00609.x}.

\bibitem[Hauser et~al.(2019)Hauser, Hilbe, Chatterjee, and Nowak]{Hauser2019SocialUnequals}
Oliver~P. Hauser, Christian Hilbe, Krishnendu Chatterjee, and Martin~A. Nowak.
\newblock {Social dilemmas among unequals}.
\newblock \emph{Nature}, 572\penalty0 (7770):\penalty0 524--527, 8 2019.
\newblock ISSN 14764687.
\newblock \doi{10.1038/s41586-019-1488-5}.

\bibitem[He et~al.(2016)He, Boyd-Graber, Kwok, and Daume]{He2016}
He~He, Jordan Boyd-Graber, Kevin Kwok, and Hal Daume.
\newblock {Opponent modeling in deep reinforcement learning}.
\newblock \emph{33rd International Conference on Machine Learning, ICML 2016}, 4:\penalty0 2675--2684, 2016.

\bibitem[Huang and Zhou(2022)]{Huang2022Importance-AwareLearning}
Xiufeng Huang and Sheng Zhou.
\newblock {Importance-Aware Message Exchange and Prediction for Multi-Agent Reinforcement Learning}.
\newblock \emph{2022 IEEE Global Communications Conference, GLOBECOM 2022 - Proceedings}, pages 6493--6498, 2022.
\newblock \doi{10.1109/GLOBECOM48099.2022.10001408}.

\bibitem[Hughes et~al.(2018)Hughes, Leibo, Phillips, and Tuyls]{Hughes2018InequityDilemmas}
Edward Hughes, Joel~Z. Leibo, Matthew Phillips, and Karl Tuyls.
\newblock {Inequity aversion improves cooperation in intertemporal social dilemmas}.
\newblock \emph{Advances in Neural Information Processing Systems}, pages 3326--3336, 2018.
\newblock ISSN 10495258.

\bibitem[Iqbal and Sha(2019)]{Iqbal2019Actor-attention-criticLearning}
Shariq Iqbal and Fei Sha.
\newblock {Actor-attention-critic for multi-agent reinforcement learning}.
\newblock \emph{36th International Conference on Machine Learning, ICML 2019}, 2019-June:\penalty0 5261--5270, 2019.

\bibitem[Jaques et~al.(2019)Jaques, Lazaridou, Hughes, Gulcehre, Ortega, Strouse, Leibo, and de~Freitas]{Jaques2019SocialLearning}
Natasha Jaques, Angeliki Lazaridou, Edward Hughes, Caglar Gulcehre, Pedro~A. Ortega, D.~J. Strouse, Joel~Z. Leibo, and Nando de~Freitas.
\newblock {Social influence as intrinsic motivation for multi-agent deep reinforcement learning}.
\newblock \emph{36th International Conference on Machine Learning, ICML 2019}, 2019-June:\penalty0 5372--5381, 2019.

\bibitem[Jiang and Lu(2022)]{Jiang2022I2QAlgorithm}
Jiechuan Jiang and Zongqing Lu.
\newblock {I2Q : A Fully Decentralized Q-Learning Algorithm}.
\newblock \emph{Advances in Neural Information Processing Systems}, 35:\penalty0 20469--20481, 2022.

\bibitem[Jin et~al.(2021)Jin, Wei, Yuan, and Zhang]{Jin2021HierarchicalControl}
Yue Jin, Shuangqing Wei, Jian Yuan, and Xudong Zhang.
\newblock {Hierarchical and Stable Multiagent Reinforcement Learning for Cooperative Navigation Control}.
\newblock \emph{IEEE Transactions on Neural Networks and Learning Systems}, 34\penalty0 (1):\penalty0 90--103, 2021.
\newblock ISSN 21622388.
\newblock \doi{10.1109/TNNLS.2021.3089834}.

\bibitem[Kim et~al.(2021)Kim, Park, and Sung]{Kim2021CommunicationSharing}
Woojun Kim, Jongeui Park, and Youngchul Sung.
\newblock {Communication in Multi-Agent Reinforcement Learning: Intention Sharing}.
\newblock \emph{ICLR}, pages 1--15, 2021.

\bibitem[Kollock(1998)]{Kollock1998SOCIALCooperation}
Peter Kollock.
\newblock {SOCIAL DILEMMAS: The Anatomy of Cooperation}.
\newblock Technical report, 1998.
\newblock URL \url{www.sscnet.ucla.edu/soc/faculty/kollock/dilemmas}.

\bibitem[Krouka et~al.(2022)Krouka, Elgabli, Issaid, and Bennis]{Krouka2022Communication-EfficientLearning}
Mounssif Krouka, Anis Elgabli, Chaouki~Ben Issaid, and Mehdi Bennis.
\newblock {Communication-Efficient and Federated Multi-Agent Reinforcement Learning}.
\newblock \emph{IEEE Transactions on Cognitive Communications and Networking}, 8\penalty0 (1):\penalty0 311--320, 2022.
\newblock ISSN 23327731.
\newblock \doi{10.1109/TCCN.2021.3130993}.

\bibitem[Kuba et~al.(2022)Kuba, Chen, Wen, Wen, Sun, Wang, and Yang]{Kuba2022TrustLearning}
Jakub~Grudzien Kuba, Ruiqing Chen, Muning Wen, Ying Wen, Fanglei Sun, Jun Wang, and Yaodong Yang.
\newblock {Trust Region Policy Optimisation in Multi-Agent Reinforcement Learning}.
\newblock \emph{International Conference on Learning Representations}, page 1046, 2022.

\bibitem[Lei et~al.(2022)Lei, Ye, Xiao, Skoglund, and Han]{Lei2022AdaptiveIoT}
Wanlu Lei, Yu~Ye, Ming Xiao, Mikael Skoglund, and Zhu Han.
\newblock {Adaptive Stochastic ADMM for Decentralized Reinforcement Learning in Edge IoT}.
\newblock \emph{IEEE Internet of Things Journal}, 9\penalty0 (22):\penalty0 22958--22971, 2022.
\newblock ISSN 23274662.
\newblock \doi{10.1109/JIOT.2022.3187067}.

\bibitem[Leibo et~al.(2017)Leibo, Zambaldi, Lanctot, Marecki, and Graepel]{Leibo2017Multi-agentDilemmas}
Joel~Z. Leibo, Vinicius Zambaldi, Marc Lanctot, Janusz Marecki, and Thore Graepel.
\newblock {Multi-agent Reinforcement Learning in Sequential Social Dilemmas}.
\newblock \emph{Proceedings of the 16th International Conference on Autonomous Agents and Multiagent Systems}, pages 464--473, 2017.
\newblock URL \url{http://arxiv.org/abs/1702.03037}.

\bibitem[Lowe et~al.(2017)Lowe, Wu, Tamar, Harb, Abbeel, and Mordatch]{Lowe2017Multi-agentEnvironments}
Ryan Lowe, Yi~Wu, Aviv Tamar, Jean Harb, Pieter Abbeel, and Igor Mordatch.
\newblock {Multi-agent actor-critic for mixed cooperative-competitive environments}.
\newblock \emph{Advances in Neural Information Processing Systems}, 2017-Decem:\penalty0 6380--6391, 2017.
\newblock ISSN 10495258.

\bibitem[Macy and Flache()]{MacyLearningDilemmas}
Michael~W Macy and Andreas Flache.
\newblock {Learning dynamics in social dilemmas}.
\newblock Technical report.
\newblock URL \url{www.pnas.orgcgidoi10.1073pnas.092080099}.

\bibitem[Milinski et~al.(2002)Milinski, Semmann, and Krambeck]{Milinski2002ReputationCommons}
M.~Milinski, D.~Semmann, and HJ. Krambeck.
\newblock {Reputation helps solve the `tragedy of the commons'}.
\newblock \emph{Nature}, 415\penalty0 (6870):\penalty0 424--426, 1 2002.
\newblock ISSN 00368075.
\newblock \doi{10.1126/science.1064748}.

\bibitem[Omidshafiei et~al.(2017)Omidshafiei, Pazis, Amato, How, and Vian]{Omidshafiei2017DeepObservability}
Shayegan Omidshafiei, Jason Pazis, Christopher Amato, Jonathan~P How, and John Vian.
\newblock {Deep Decentralized Multi-task Multi-Agent Reinforcement Learning under Partial Observability}.
\newblock 2017.
\newblock \doi{10.5555/3305890.3305958}.

\bibitem[Ostrom(1990)]{Ostrom1990GoverningAction}
E.~Ostrom.
\newblock \emph{{Governing the commons: the evolution of institutions for collective action}}, volume~32.
\newblock 1990.
\newblock ISBN 0521371015.
\newblock \doi{10.2307/3146384}.

\bibitem[Peng et~al.(2021)Peng, Rashid, Schroeder~de Witt, Kamienny, Torr, B{\"{o}}hmer, and Whiteson]{Peng2021FACMAC:Gradients}
Bei Peng, Tabish Rashid, Christian~A. Schroeder~de Witt, Pierre~Alexandre Kamienny, Philip~H.S. Torr, Wendelin B{\"{o}}hmer, and Shimon Whiteson.
\newblock {FACMAC: Factored Multi-Agent Centralised Policy Gradients}.
\newblock \emph{Advances in Neural Information Processing Systems}, 15\penalty0 (NeurIPS):\penalty0 12208--12221, 2021.
\newblock ISSN 10495258.

\bibitem[Qiu et~al.(2023)Qiu, Jin, Yu, Wang, Wang, and Zhang]{Qiu2023ImprovingControl}
Yunbo Qiu, Yue Jin, Lebin Yu, Jian Wang, Yu~Wang, and Xudong Zhang.
\newblock {Improving Sample Efficiency of Multi-Agent Reinforcement Learning with Non-expert Policy for Flocking Control}.
\newblock \emph{IEEE Internet of Things Journal}, 10\penalty0 (14):\penalty0 14014--14027, 2023.
\newblock \doi{10.1109/JIOT.2023.3240671}.

\bibitem[Schulman et~al.(2015)Schulman, Levine, Moritz, Jordan, and Abbeel]{Schulman2015TrustOptimization}
John Schulman, Sergey Levine, Philipp Moritz, Michael Jordan, and Pieter Abbeel.
\newblock {Trust region policy optimization}.
\newblock \emph{32nd International Conference on Machine Learning, ICML 2015}, 3:\penalty0 1889--1897, 2015.

\bibitem[Schulman et~al.(2016)Schulman, Moritz, Levine, Jordan, and Abbeel]{Schulman2016High-dimensionalEstimation}
John Schulman, Philipp Moritz, Sergey Levine, Michael~I. Jordan, and Pieter Abbeel.
\newblock {High-dimensional continuous control using generalized advantage estimation}.
\newblock \emph{4th International Conference on Learning Representations, ICLR 2016 - Conference Track Proceedings}, pages 1--14, 2016.

\bibitem[Schulman et~al.(2017)Schulman, Wolski, Dhariwal, Radford, and Klimov]{Schulman2017ProximalAlgorithms}
John Schulman, Filip Wolski, Prafulla Dhariwal, Alec Radford, and Oleg Klimov.
\newblock {Proximal Policy Optimization Algorithms}.
\newblock \emph{arXiv preprint arXiv:1707.06347}, 2017.

\bibitem[Sha et~al.(2021)Sha, Zhang, and You]{Sha2021PolicyNetworks}
Xingyu Sha, Jiaqi Zhang, and Keyou You.
\newblock {Policy evaluation for reinforcement learning over asynchronous multi-agent networks}.
\newblock \emph{Chinese Control Conference, CCC}, 2021-July:\penalty0 5373--5378, 2021.
\newblock ISSN 21612927.
\newblock \doi{10.23919/CCC52363.2021.9550466}.

\bibitem[Siedler and Alpha()]{SiedlerDynamicReforestation}
Philipp~D Siedler and Aleph Alpha.
\newblock {Dynamic Collaborative Multi-Agent Reinforcement Learning Communication for Autonomous Drone Reforestation}.
\newblock \penalty0 (NeurIPS 2022).

\bibitem[Stankovic et~al.(2022{\natexlab{a}})Stankovic, Beko, and Stankovic]{Stankovic2022DistributedWeightings}
Milos~S. Stankovic, Marko Beko, and Srdjan~S. Stankovic.
\newblock {Distributed Actor-Critic Learning Using Emphatic Weightings}.
\newblock \emph{2022 8th International Conference on Control, Decision and Information Technologies, CoDIT 2022}, pages 1167--1172, 2022{\natexlab{a}}.
\newblock \doi{10.1109/CoDIT55151.2022.9804022}.

\bibitem[Stankovic et~al.(2022{\natexlab{b}})Stankovic, Beko, and Stankovic]{Stankovic2022ConvergentDifference}
Miloš~S. Stankovic, Marko Beko, and Srdjan~S. Stankovic.
\newblock {Convergent Distributed Actor-Critic Algorithm Based on Gradient Temporal Difference}.
\newblock \emph{European Signal Processing Conference}, 2022-Augus:\penalty0 2066--2070, 2022{\natexlab{b}}.
\newblock ISSN 22195491.
\newblock \doi{10.23919/eusipco55093.2022.9909762}.

\bibitem[Su and Lu(2022)]{Su2022DecentralizedOptimization}
Kefan Su and Zongqing Lu.
\newblock {Decentralized Policy Optimization}.
\newblock \emph{arXiv preprint arXiv:2211.03032}, 2022.

\bibitem[Sun et~al.(2020)Sun, Shen, and How]{Sun2020ScalingGraph}
Chuangchuang Sun, Macheng Shen, and Jonathan~P. How.
\newblock {Scaling up multiagent reinforcement learning for robotic systems: Learn an adaptive sparse communication graph}.
\newblock \emph{IEEE International Conference on Intelligent Robots and Systems}, pages 11755--11762, 2020.
\newblock ISSN 21530866.
\newblock \doi{10.1109/IROS45743.2020.9341303}.

\bibitem[Sun et~al.(2022)Sun, Devlin, Beck, Hofmann, and Whiteson]{Sun2022TrustNon-stationarity}
Mingfei Sun, Sam Devlin, Jacob Beck, Katja Hofmann, and Shimon Whiteson.
\newblock {Trust Region Bounds for Decentralized PPO Under Non-stationarity}.
\newblock \emph{Proceedings of the 2023 International Conference on Autonomous Agents and Multiagent Systems}, pages 5--13, 2022.
\newblock URL \url{http://arxiv.org/abs/2202.00082}.

\bibitem[Suttle et~al.(2020)Suttle, Yang, Zhang, Wang, Basar, and Liu]{Suttle2020ALearning}
Wesley Suttle, Zhuoran Yang, Kaiqing Zhang, Zhaoran Wang, Tamer Basar, and Ji~Liu.
\newblock {A multi-agent off-policy actor-critic algorithm for distributed reinforcement learning}.
\newblock \emph{IFAC-PapersOnLine}, 53:\penalty0 1549--1554, 2020.
\newblock ISSN 24058963.
\newblock \doi{10.1016/j.ifacol.2020.12.2021}.

\bibitem[Tennant et~al.(2023)Tennant, Hailes, and Musolesi]{Tennant2023ModelingLearning}
Elizaveta Tennant, Stephen Hailes, and Mirco Musolesi.
\newblock {Modeling Moral Choices in Social Dilemmas with Multi-Agent Reinforcement Learning}.
\newblock \emph{arXiv preprint arXiv:2301.08491}, 2023.
\newblock URL \url{https://arxiv.org/abs/2301.08491v1}.

\bibitem[Van~Lange et~al.(2013)Van~Lange, Joireman, Parks, and Van~Dijk]{VanLange2013TheReview}
Paul~A.M. Van~Lange, Jeff Joireman, Craig~D. Parks, and Eric Van~Dijk.
\newblock {The psychology of social dilemmas: A review}.
\newblock \emph{Organizational Behavior and Human Decision Processes}, 120\penalty0 (2):\penalty0 125--141, 3 2013.
\newblock ISSN 07495978.
\newblock \doi{10.1016/j.obhdp.2012.11.003}.

\bibitem[Wang et~al.(2022)Wang, Damani, Wang, Cao, and Sartoretti]{Wang2022DistributedReview}
Yutong Wang, Mehul Damani, Pamela Wang, Yuhong Cao, and Guillaume Sartoretti.
\newblock {Distributed Reinforcement Learning for Robot Teams: A Review}.
\newblock 2022.
\newblock URL \url{http://arxiv.org/abs/2204.03516}.

\bibitem[Wen et~al.(2019)Wen, Yang, Luo, Wang, and Pan]{Wen2019}
Ying Wen, Yaodong Yang, Rui Luo, Jun Wang, and Wei Pan.
\newblock {Probabilistic recursive reasoning for multi-agent reinforcement learning}.
\newblock \emph{7th International Conference on Learning Representations, ICLR 2019}, pages 1--20, 2019.

\bibitem[Wu et~al.(2021)Wu, Yu, Ye, Zhang, Piao, and Zhuo]{Wu2021CoordinatedOptimization}
Zifan Wu, Chao Yu, Deheng Ye, Junge Zhang, Haiyin Piao, and Hankz~Hankui Zhuo.
\newblock {Coordinated Proximal Policy Optimization}.
\newblock \emph{Advances in Neural Information Processing Systems}, 32:\penalty0 26437--26448, 2021.
\newblock ISSN 10495258.

\bibitem[Xia et~al.(2022)Xia, Du, Wang, Jiang, Ren, Li, and Han]{Xia2022Multi-AgentTracking}
Zhaoyue Xia, Jun Du, Jingjing Wang, Chunxiao Jiang, Yong Ren, Gang Li, and Zhu Han.
\newblock {Multi-Agent Reinforcement Learning Aided Intelligent UAV Swarm for Target Tracking}.
\newblock \emph{IEEE Transactions on Vehicular Technology}, 71\penalty0 (1):\penalty0 931--945, 2022.
\newblock ISSN 19399359.
\newblock \doi{10.1109/TVT.2021.3129504}.

\bibitem[Yi et~al.(2022)Yi, Li, Wang, and Lu]{Yi2022LearningLearning}
Yuxuan Yi, Ge~Li, Yaowei Wang, and Zongqing Lu.
\newblock {Learning to Share in Multi-Agent Reinforcement Learning}.
\newblock \emph{ICLR 2022 Workshop on Gamification and Multiagent Solutions}, 2022.
\newblock URL \url{http://arxiv.org/abs/2112.08702}.

\bibitem[Zhang et~al.(2018{\natexlab{a}})Zhang, Yang, and Basar]{Zhang2018NetworkedSpaces}
Kaiqing Zhang, Zhuoran Yang, and Tamer Basar.
\newblock {Networked Multi-Agent Reinforcement Learning in Continuous Spaces}.
\newblock \emph{Proceedings of the IEEE Conference on Decision and Control}, 2018-Decem\penalty0 (Cdc):\penalty0 2771--2776, 2018{\natexlab{a}}.
\newblock ISSN 25762370.
\newblock \doi{10.1109/CDC.2018.8619581}.

\bibitem[Zhang et~al.(2018{\natexlab{b}})Zhang, Yang, Liu, Zhang, and Ba{\c{s}}ar]{Zhang2018FullyAgents}
Kaiqing Zhang, Zhuoran Yang, Han Liu, Tong Zhang, and Tamer Ba{\c{s}}ar.
\newblock {Fully decentralized multi-agent reinforcement learning with networked agents}.
\newblock \emph{35th International Conference on Machine Learning, ICML 2018}, 13:\penalty0 9340--9371, 2018{\natexlab{b}}.

\bibitem[Zhang et~al.(2020)Zhang, Yang, Liu, Zhang, and Basar]{Zhang2020Finite-sampleData}
Kaiqing Zhang, Zhuoran Yang, Han Liu, Tong Zhang, and Tamer Basar.
\newblock {Finite-sample analysis for decentralized cooperative multi-agent reinforcement learning from batch data}.
\newblock \emph{IFAC-PapersOnLine}, 53\penalty0 (2):\penalty0 1049--1056, 2020.
\newblock ISSN 24058963.
\newblock \doi{10.1016/j.ifacol.2020.12.1290}.

\bibitem[Zhang and Zavlanos(2019)]{Zhang2019DistributedConsensus}
Yan Zhang and Michael~M. Zavlanos.
\newblock {Distributed off-Policy Actor-Critic Reinforcement Learning with Policy Consensus}.
\newblock \emph{Proceedings of the IEEE Conference on Decision and Control}, 2019-Decem\penalty0 (Cdc):\penalty0 4674--4679, 2019.
\newblock ISSN 25762370.
\newblock \doi{10.1109/CDC40024.2019.9029969}.

\bibitem[Zhao et~al.(2020)Zhao, Yi, and Li]{Zhao2020DistributedLearning}
Xiaoxiao Zhao, Peng Yi, and Li~Li.
\newblock {Distributed policy evaluation via inexact ADMM in multi-agent reinforcement learning}.
\newblock \emph{Control Theory and Technology}, 18\penalty0 (4):\penalty0 362--378, 2020.
\newblock ISSN 21980942.
\newblock \doi{10.1007/s11768-020-00007-x}.

\bibitem[Zheng et~al.(2018)Zheng, Meng, Hao, Zhang, Yang, and Fan]{Zheng2018}
Yan Zheng, Zhaopeng Meng, Jianye Hao, Zongzhang Zhang, Tianpei Yang, and Changjie Fan.
\newblock {A deep Bayesian policy reuse approach against non-stationary agents}.
\newblock \emph{Advances in Neural Information Processing Systems}, 2018-Decem\penalty0 (NeurIPS):\penalty0 954--964, 2018.
\newblock ISSN 10495258.

\end{thebibliography}

\appendix
\section{Appendix}
\subsection{Proofs}
\subsubsection{Proof of Lemma \ref{theorem_1}} \label{proof_1}
\textbf{Lemma \ref{theorem_1}}
The following bound holds for the difference between the expected returns of the current policy $\boldsymbol{\pi}_{old}$ and another policy $\boldsymbol{\pi}_{new}$
\begin{equation}\label{TRPO}
    \begin{aligned}
        \eta(\boldsymbol{\pi}_{new}) \geq \eta(\boldsymbol{\pi}_{old}) + \zeta_{\boldsymbol{\pi}_{old}}(\boldsymbol{\pi}_{new}) - C \cdot D_{KL}^{max}(\boldsymbol{\pi}_{old} || \boldsymbol{\pi}_{new}),
    \end{aligned}
\end{equation}
where 
\begin{equation}
\begin{aligned}
     & \zeta_{\boldsymbol{\pi}_{old}}(\boldsymbol{\pi}_{new}) = \mathbb{E}_{s \sim d^{\boldsymbol{\pi}_{old}}(s), \boldsymbol{a}\sim \boldsymbol{\pi}_{new}(\cdot|s)} \left[\sum_{i} A_i^{\boldsymbol{\pi}_{old}}(s, \boldsymbol{a})\right],   \\
     & C = \frac{4 \max_{s, \boldsymbol{a}}|\sum_{i} A_i^{\boldsymbol{\pi}_{old}}(s, \boldsymbol{a})|\gamma}{(1-\gamma)^2}\\
     & D_{KL}^{max}(\boldsymbol{\pi}_{old} || \boldsymbol{\pi}_{new}) = \max_{s}D_{KL}(\boldsymbol{\pi}_{old}(\cdot|s) || \boldsymbol{\pi}_{new}(\cdot|s)).
\end{aligned}
\end{equation}

\begin{lemma}
Given two joint policies $\boldsymbol{\pi}_{old}$ and $\boldsymbol{\pi}_{new}$, 
\begin{equation} \label{lemma_1}
    \eta(\boldsymbol{\pi}_{new}) =  \eta(\boldsymbol{\pi}_{old}) + \mathbb{E}_{\tau \sim  \boldsymbol{\pi}_{new}} \left[\sum_{i=1}^N \sum_{t=0}^\infty \gamma^t A_i^{\boldsymbol{\pi}_{old}}(s_t, \boldsymbol{a}_t)\right], 
\end{equation}
\end{lemma}
where $\mathbb{E}_{\tau \sim \boldsymbol{\pi}_{new}}[\cdot]$ means the expectation is computed over trajectories where the initial state distribution $s_0 \sim d(s_0)$, action selection $\boldsymbol{a}_t \sim \boldsymbol{\pi}_{new}(\cdot| s_t)$, and state transitions $s_{t+1} \sim \mathcal{P}(\cdot| s_t, \boldsymbol{a}_t)$.

\emph{Proof:}
The expected discounted reward of the joint policy, i.e., Eq.~\ref{objective}, can be expressed as 
\begin{equation}
    \eta(\boldsymbol{\pi})= \sum_{i=1}^N \mathbb{E}_{s_0 \sim d(s_0)} \left[V_i^{\boldsymbol{\pi}} (s_0)\right].
\end{equation}
Using $A_i^{\boldsymbol{\pi}_{old}}(s_t, \boldsymbol{a}_t) = \mathbb{E}_{s'} [r_t^i + \gamma V_i^{\boldsymbol{\pi}_{old}}(s') - V_i^{\boldsymbol{\pi}_{old}}(s)]$, we have
\begin{equation}
\begin{aligned}
    &\mathbb{E}_{\tau \sim \boldsymbol{\pi}_{new}} \left[\sum_{i=1}^N \sum_{t=0}^\infty \gamma^t A_i^{\boldsymbol{\pi}_{old}}(s_t, \boldsymbol{a}_t)\right] \\
    & = 
    {\mathbb{E}_{\tau \sim \boldsymbol{\pi}_{new}} \left[\sum_{i=1}^N \sum_{t=0}^\infty \gamma^t (r_t^i + \gamma V_i^{\boldsymbol{\pi}_{old}}(s_{t+1}) - V_i^{\boldsymbol{\pi}_{old}}(s_t))\right]} \\
    & = {\mathbb{E}_{\tau \sim \boldsymbol{\pi}_{new}} \left[\sum_{i=1}^N \sum_{t=0}^\infty \gamma^{t+1} V_i^{\boldsymbol{\pi}_{old}}(s_{t+1}) - \sum_{t=0}^\infty \gamma^t V_i^{\boldsymbol{\pi}_{old}}(s_t) + \sum_{t=0}^\infty \gamma^t r_t^i \right]}  \\
    & = {\mathbb{E}_{\tau \sim \boldsymbol{\pi}_{new}} \left[\sum_{i=1}^N \sum_{\bf{{t=1}}}^\infty \gamma^t V_i^{\boldsymbol{\pi}_{old}}(s_{t}) - \sum_{t=0}^\infty \gamma^t V_i^{\boldsymbol{\pi}_{old}}(s_t) + \sum_{t=0}^\infty \gamma^t r_t^i \right]}  \\
    & = \mathbb{E}_{\tau \sim \boldsymbol{\pi}_{new}} \left[\sum_{i=1}^N (-V_i^{\boldsymbol{\pi}_{old}}(s_0) + \sum_{t=0}^\infty \gamma^t r_t^i) \right] \\
    & = - \sum_{i=1}^N \mathbb{E}_{s_0 \sim d(s_0)}[V_i^{\boldsymbol{\pi}_{old}}(s_0)] + \sum_{i=1}^N \mathbb{E}_{\tau \sim \boldsymbol{\pi}_{new}} \left[\sum_{t=0}^\infty \gamma^t r_t^i \right] \\
    & = - \eta(\boldsymbol{\pi}_{old}) + \eta(\boldsymbol{\pi}_{new}).
\end{aligned}    
\end{equation}
Thus, we have Eq.~\ref{lemma_1}.

Define an expected joint advantage $\bar{A}_{joint}$ as 
\begin{equation}
    \bar{A}_{joint}(s) = \mathbb{E}_{\boldsymbol{a} \sim \boldsymbol{\pi}_{new}(\cdot|s)} \left[\sum_{i=1}^N A_i^{\boldsymbol{\pi}_{old}}(s, \boldsymbol{a}) \right].
\end{equation}

Define $L_{\boldsymbol{\pi}_{old}}(\boldsymbol{\pi}_{new})$ as
\begin{equation} \label{L_old_new}
\begin{aligned}
    L_{\boldsymbol{\pi}_{old}}(\boldsymbol{\pi}_{new}) 
    & = \eta(\boldsymbol{\pi}_{old}) + \mathbb{E}_{\tau \sim \pi_{old}} \left[ \sum_{t=0}^\infty \gamma^t \bar{A}_{joint}(s_t) \right] \\
    & = \eta(\boldsymbol{\pi}_{old}) + \sum_s \sum_{t=0}^\infty \gamma^t P(s_t = s| \boldsymbol{\pi}_{old}) \bar{A}_{joint}(s) .
\end{aligned}
\end{equation}

Leveraging the Lemma 2, Lemma 3, and Theorem 1 provided by TRPO \citep{Schulman2015TrustOptimization}, we have  
\begin{equation}
    \left|\eta(\boldsymbol{\pi}_{new}) - L_{\boldsymbol{\pi}_{old}}(\boldsymbol{\pi}_{new}) \right| \leq C \cdot (\max_{s}D_{TV}(\boldsymbol{\pi}_{old}(\cdot|s) || \boldsymbol{\pi}_{new}(\cdot|s)))^2.
\end{equation}

Based on the relationship: $(D_{TV}(p||q))^2 \leq D_{KL}(q||q)$, we have 
\begin{equation} \label{ineq}
    \left|\eta(\boldsymbol{\pi}_{new}) - L_{\boldsymbol{\pi}_{old}}(\boldsymbol{\pi}_{new}) \right| \leq C \cdot D_{KL}^{max}(\boldsymbol{\pi}_{old} || \boldsymbol{\pi}_{new}).
\end{equation}

For the second term of the RHS of Eq.~\ref{L_old_new}, we have the following equivalent form

\begin{equation}
\begin{aligned}
    & \sum_s \sum_{t=0}^\infty \gamma^t P(s_t = s| \boldsymbol{\pi}_{old}) \bar{A}_{joint}(s) \\
    & = \sum_s \sum_{t=0}^\infty \gamma^t  P(s_t = s| \boldsymbol{\pi}_{old}) \bar{A}_{joint}(s)  \\
    & = \sum_s d^{\boldsymbol{\pi}_{old}}(s) \bar{A}_{joint}(s) \\
    & = \sum_s d^{\boldsymbol{\pi}_{old}}(s) \mathbb{E}_{\boldsymbol{a} \sim \boldsymbol{\pi}_{new}(\cdot|s)}\left[\sum_{i=1}^N A_i^{\boldsymbol{\pi}_{old}}(s, \boldsymbol{a}) \right] \\
    & = \zeta_{\boldsymbol{\pi}_{old}}(\boldsymbol{\pi}_{new}),
\end{aligned}
\end{equation}
where $d^{\boldsymbol{\pi}}$ denotes the state visitation distribution under policy $\boldsymbol{\pi}$, and the third line is derived based on the property $d^{\boldsymbol{\pi}_{old}}(s) = P(s_0=s) +\gamma P(s_1=s) + \gamma^2 P(s_2=s) + \cdots$.

Thus, we have $L_{\boldsymbol{\pi}_{old}}(\boldsymbol{\pi}_{new}) = \eta(\boldsymbol{\pi}_{old}) + \zeta_{\boldsymbol{\pi}_{old}}(\boldsymbol{\pi}_{new})$. Then, replacing $L_{\boldsymbol{\pi}_{old}}(\boldsymbol{\pi}_{new})$ in Eq.~\ref{ineq}, we have  
\begin{equation}
    \left|\eta(\boldsymbol{\pi}_{new}) - (\eta(\boldsymbol{\pi}_{old}) + \zeta_{\boldsymbol{\pi}_{old}}(\boldsymbol{\pi}_{new})) \right| \leq C \cdot D_{KL}^{max}(\boldsymbol{\pi}_{old} || \boldsymbol{\pi}_{new}),
\end{equation}
and thus Lemma \ref{theorem_1} is proved.

\subsubsection{Proof of Lemma \ref{theorem_2}} \label{proof_2}
\textbf{Lemma \ref{theorem_2}}
\emph{The discrepancy between $\zeta_{\boldsymbol{\pi}'}(\boldsymbol{\tilde{\Pi}})$ and the sum of the expected individual advantages calculated with policy $\boldsymbol{\pi}'$ over the true joint policy $\boldsymbol{\pi}$, i.e., $\zeta_{\boldsymbol{\pi}'}(\boldsymbol{\pi})$, is upper bounded as follows.}
\begin{equation} \label{bound}
     \zeta_{\boldsymbol{\pi}'}(\boldsymbol{\tilde{\Pi}}) - \zeta_{\boldsymbol{\pi}'}(\boldsymbol{\pi}) \leq f^{\boldsymbol{\pi}'} + \sum_{i} \frac{1}{2} \max_{s, \boldsymbol{a}} \left| A_i^{\boldsymbol{\pi}'}(s, \boldsymbol{a}) \right|  \cdot \sum_{s, \boldsymbol{a}}\left(\boldsymbol{\tilde{\pi}}^i(\boldsymbol{a}|s) - \boldsymbol{\pi}(\boldsymbol{a}|s)\right)^2 , 
\end{equation}
where 
\begin{equation} \label{f_pi}
    f^{\boldsymbol{\pi}'} = \sum_{i} \frac{1}{2} \max_{s, \boldsymbol{a}} \left| A_i^{\boldsymbol{\pi}'}(s, \boldsymbol{a}) \right|  \cdot |\mathcal{A}| \cdot \Vert d^{\boldsymbol{\pi}'} \Vert_2^2, 
\end{equation}
and $\|d^{\pi'}\|_2^2 = \sum_s (d^{\pi'}(s))^2$.

\emph{Proof:}
    \begin{equation} 
    \begin{aligned}
     \zeta_{\boldsymbol{\pi}'}(\boldsymbol{\tilde{\Pi}}) - \zeta_{\boldsymbol{\pi}'}(\boldsymbol{\pi}) &= \sum_{i} \mathbb{E}_{s \sim d^{\boldsymbol{\pi}'}(s), \boldsymbol{a}\sim \boldsymbol{\tilde{\pi}}^i(\boldsymbol{a}|s)} \left[A_i^{\boldsymbol{\pi}'}(s, \boldsymbol{a})\right] - \mathbb{E}_{s \sim d^{\boldsymbol{\pi}'}(s), \boldsymbol{a}\sim \boldsymbol{\pi}(\boldsymbol{a}|s)} \left[ A_i^{\boldsymbol{\pi}'}(s, \boldsymbol{a}) \right] \\
     & = \sum_{i} \sum_{s, \boldsymbol{a}} d^{\boldsymbol{\pi}'}(s) (\boldsymbol{\tilde{\pi}}^i(a|s) - \boldsymbol{\pi}(a|s)) A_i^{\boldsymbol{\pi}'}(s, a),\\  
     & \leq \sum_{i} \max_{s, \boldsymbol{a}} \left| A_i^{\boldsymbol{\pi}'}(s, \boldsymbol{a}) \right|  \cdot \left| \sum_{s, \boldsymbol{a}} d^{\boldsymbol{\pi}'}(s) \left(\boldsymbol{\tilde{\pi}}^i(\boldsymbol{a}|s) - \boldsymbol{\pi}(\boldsymbol{a}|s)\right) \right| \\
     & \leq \sum_{i} \max_{s, \boldsymbol{a}} \left| A_i^{\boldsymbol{\pi}'}(s, \boldsymbol{a}) \right|  \cdot \sum_{s, \boldsymbol{a}} \frac{1}{2} \left(d^{\boldsymbol{\pi}'}(s)^2 +\left(\boldsymbol{\tilde{\pi}}^i(\boldsymbol{a}|s) - \boldsymbol{\pi}(\boldsymbol{a}|s)\right)^2 \right) \\
     & = \sum_{i} \frac{1}{2} \max_{s, \boldsymbol{a}} \left| A_i^{\boldsymbol{\pi}'}(s, \boldsymbol{a}) \right|  \cdot \sum_{s, \boldsymbol{a}} \left( d^{\boldsymbol{\pi}'}(s)^2 +\left(\boldsymbol{\tilde{\pi}}^i(\boldsymbol{a}|s) - \boldsymbol{\pi}(\boldsymbol{a}|s)\right)^2 \right) \\
     & = \sum_{i} \frac{1}{2} \max_{s, \boldsymbol{a}} \left| A_i^{\boldsymbol{\pi}'}(s, \boldsymbol{a}) \right|  \cdot \left(|\mathcal{A}| \cdot \Vert d^{\boldsymbol{\pi}'} \Vert_2^2 + \sum_{s, \boldsymbol{a}} \left(\boldsymbol{\tilde{\pi}}^i(\boldsymbol{a}|s) - \boldsymbol{\pi}(\boldsymbol{a}|s)\right)^2 \right) \\
     & = f^{\boldsymbol{\pi}'} + \sum_{i} \frac{1}{2} \max_{s, \boldsymbol{a}} \left| A_i^{\boldsymbol{\pi}'}(s, \boldsymbol{a}) \right|  \cdot \sum_{s, \boldsymbol{a}}\left(\boldsymbol{\tilde{\pi}}^i(\boldsymbol{a}|s) - \boldsymbol{\pi}(\boldsymbol{a}|s)\right)^2
\end{aligned}
\end{equation}
where 
\begin{equation*}
    f^{\boldsymbol{\pi}'} = \sum_{i} \frac{1}{2} \max_{s, \boldsymbol{a}} \left| A_i^{\boldsymbol{\pi}'}(s, \boldsymbol{a}) \right|  \cdot |\mathcal{A}| \cdot \Vert d^{\boldsymbol{\pi}'} \Vert_2^2. 
\end{equation*}

\subsubsection{Proof of Theorem \ref{theorem_surrogate}} \label{proof_3}

\textbf{Theorem \ref{theorem_surrogate}}
\emph{
    The discrepancy between the return of the newer joint policy and the value of $\zeta_{\boldsymbol{\pi}_{old}}(\boldsymbol{\tilde{\Pi}}_{new})$ is lower bounded as follows:
    \begin{equation} \label{lower_bound_app}
    \begin{aligned}
        \eta(\boldsymbol{\pi}_{new}) - \zeta_{\boldsymbol{\pi}_{old}}(\boldsymbol{\tilde{\Pi}}_{new}) \geq  &  \eta(\boldsymbol{\pi}_{old}) - C \cdot \sum_i D_{KL}^{max}(\pi^{ii}_{old} || \pi^{ii}_{new}) - f^{\boldsymbol{\pi}_{old}} \\
        & - \sum_{i} \frac{1}{2} \max_{s, \boldsymbol{a}} \left| A_i^{\boldsymbol{\pi}_{old}}(s, \boldsymbol{a}) \right|  \cdot \sum_{s, \boldsymbol{a}}\left(\boldsymbol{\tilde{\pi}}^i_{new}(\boldsymbol{a}|s) - \boldsymbol{\pi}_{new}(\boldsymbol{a}|s)\right)^2.
    \end{aligned}
    \end{equation}
}

\emph{Proof:} 
According to Theorem~\ref{theorem_1}, we have
\begin{equation} \label{unknown_2}
     \eta(\boldsymbol{\pi}_{new}) \geq \zeta_{\boldsymbol{\pi}_{old}}(\boldsymbol{\pi}_{new}) + \eta(\boldsymbol{\pi}_{old}) - C \cdot D_{KL}^{max}(\boldsymbol{\pi}_{old} || \boldsymbol{\pi}_{new}).
\end{equation}
The KL divergence has the following property \citep{Su2022DecentralizedOptimization}:
\begin{equation} \label{unknown_3}
    \begin{aligned}
        D_{KL}^{max}(\boldsymbol{\pi}_{old} || \boldsymbol{\pi}_{new}) \leq \sum_i  D_{KL}^{max}(\pi^{ii}_{old} || \pi^{ii}_{new}).
    \end{aligned}
\end{equation}
Based on Eq.~\ref{unknown_2} and Eq.~\ref{unknown_3}, we have
\begin{equation} \label{unknown_4}
    \eta(\boldsymbol{\pi}_{new}) \geq \zeta_{\boldsymbol{\pi}_{old}}(\boldsymbol{\pi}_{new}) + \eta(\boldsymbol{\pi}_{old}) - C \cdot \sum_i  D_{KL}^{max}(\pi^{ii}_{old} || \pi^{ii}_{new}).
\end{equation}

Using Theorem~\ref{theorem_2}, $\zeta_{\boldsymbol{\pi}_{old}}(\boldsymbol{\tilde{\Pi}}_{new})$ and $\zeta_{\boldsymbol{\pi}_{old}}(\boldsymbol{\pi}_{new})$ satisfy the following inequality:
\begin{equation} \label{unknown_1}
\begin{aligned}
    & \zeta_{\boldsymbol{\pi}_{old}}(\boldsymbol{\pi}_{new}) \\
    & \geq \zeta_{\boldsymbol{\pi}_{old}}(\boldsymbol{\tilde{\Pi}}_{new}) - \sum_{i} \frac{1}{2} \max_{s, \boldsymbol{a}} \left| A_i^{\boldsymbol{\pi}_{old}}(s, \boldsymbol{a}) \right|  \cdot \sum_{s, \boldsymbol{a}} \max_s d^{\boldsymbol{\pi}_{old}}(s)^2 +(\boldsymbol{\tilde{\pi}}^i_{new}(\boldsymbol{a}|s) - \boldsymbol{\pi}_{new}(\boldsymbol{a}|s))^2.    
\end{aligned}
\end{equation}
According to Eq.~\ref{f_pi}, Eq.~\ref{unknown_1} can be transformed as:
\begin{equation} \label{unknown_1_new}
\begin{aligned}
    & \zeta_{\boldsymbol{\pi}_{old}}(\boldsymbol{\pi}_{new}) \\
    & \geq \zeta_{\boldsymbol{\pi}_{old}}(\boldsymbol{\tilde{\Pi}}_{new}) - f^{\boldsymbol{\pi}_{old}} - \sum_{i} \frac{1}{2} \max_{s, \boldsymbol{a}} \left| A_i^{\boldsymbol{\pi}_{old}}(s, \boldsymbol{a}) \right|  \cdot \sum_{s, \boldsymbol{a}}\left(\boldsymbol{\tilde{\pi}}^i_{new}(\boldsymbol{a}|s) - \boldsymbol{\pi}_{new}(\boldsymbol{a}|s)\right)^2.    
\end{aligned}
\end{equation}
By replacing $\zeta_{\boldsymbol{\pi}_{old}}(\boldsymbol{\pi}_{new})$ in Eq.~\ref{unknown_4} with the RHS of Eq.~\ref{unknown_1_new}, we can get Eq.~\ref{lower_bound_app}, and thus Theorem ~\ref{theorem_surrogate} is proved. 

\newpage
\subsection{Algorithm Illustration}
\label{algorithm_appendix}

\begin{figure}[!h]
  \centering
\includegraphics[width=1\columnwidth]{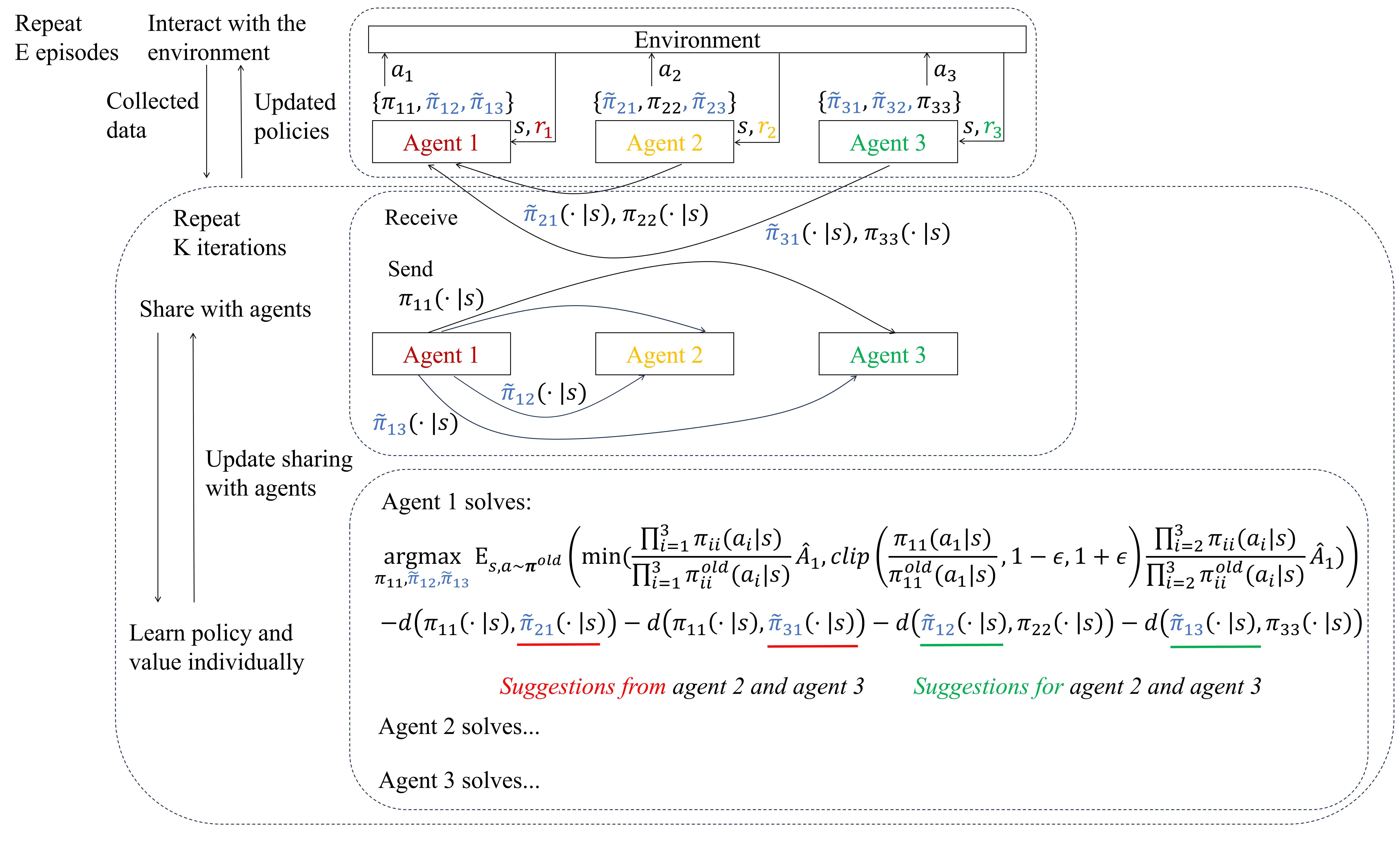}
\caption{Illustration of SS algorithm, where $d$ represents the function regarding the discrepancy term used in Eq. \ref{obj}.}
\label{illustration}
\end{figure}

Fig. \ref{illustration} shows an illustration of our SS-based MARL algorithm.

\subsection{Illustrations of Simulated Environments} 

\label{appendix_environment}

\begin{figure}[!h]
    \centering
    \begin{subfigure}[t]{0.4\textwidth}
        \centering
        \includegraphics[width=\textwidth]{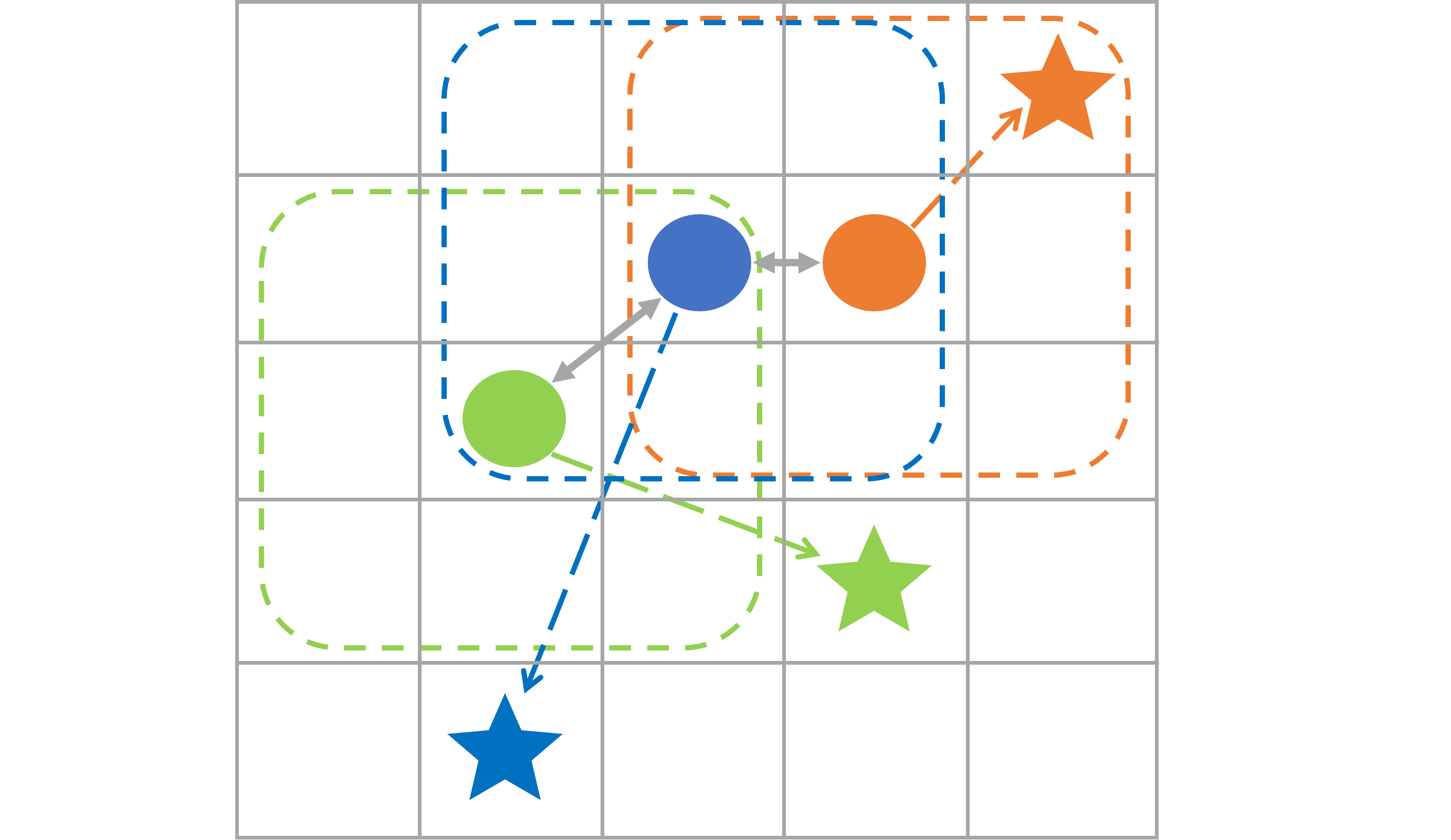}
        \caption{Cooperative navigation}
    \end{subfigure}
    \vspace{4mm} 
    \begin{subfigure}[t]{0.65\columnwidth}
        \centering
        \includegraphics[width=\textwidth]{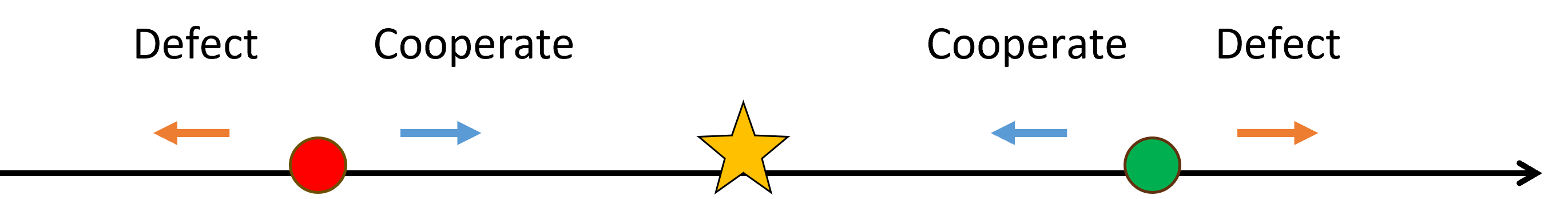}
        \caption{Cooperative predation}
    \end{subfigure}
    \caption{Illustrations of environments.}
    \label{environments}
\end{figure}

Illustrations of the Cooperative Navigation and Cooperative Predation environments are shown in Fig.~\ref{environments} (a) and (b), respectively.

\subsection{Scalability Study}

To address the scalability issue, we employ a sparse network topology and reduced communication frequency to lower computational costs. Two protocols were tested: (1) each agent randomly selected $m$ ($m \leq n$) agents for suggestion sharing, and (2) agents communicated only every two learning updates (episodes), halving the communication frequency. During communication gaps, agents updated their policies independently, omitting the last two terms in Eq.~\ref{obj} and the policy ratio $\xi_{\mathcal{N}_i}$ related to others' true policies. 

\begin{figure}[h]
    \centering
    \begin{subfigure}[t]{0.49\linewidth}
        \centering
        \includegraphics[width=\linewidth]{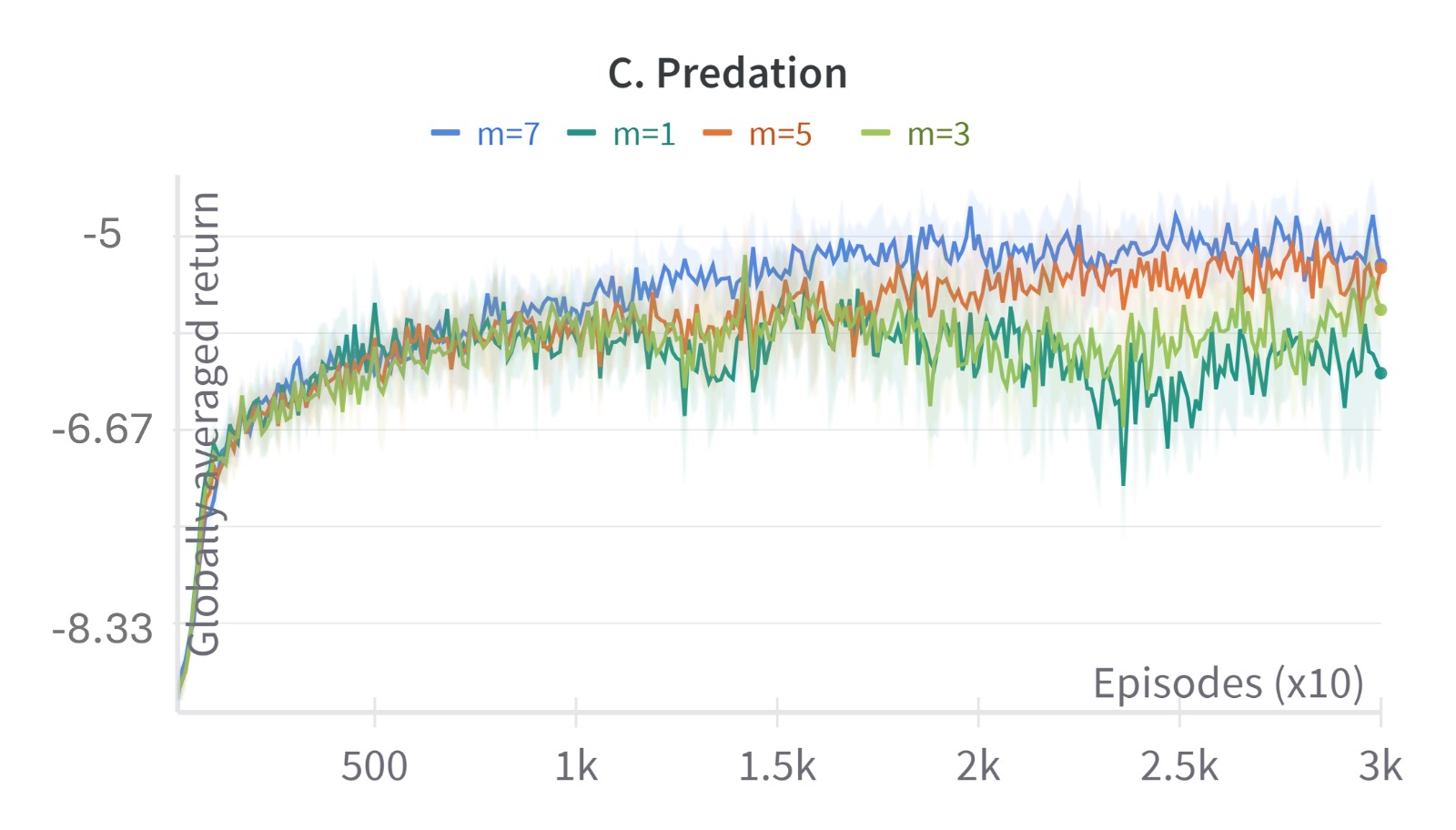}
        \caption{Reducing the number of neighbours.}
    \end{subfigure}
    \begin{subfigure}[t]{0.49\linewidth}
        \centering
        \includegraphics[width=\linewidth]{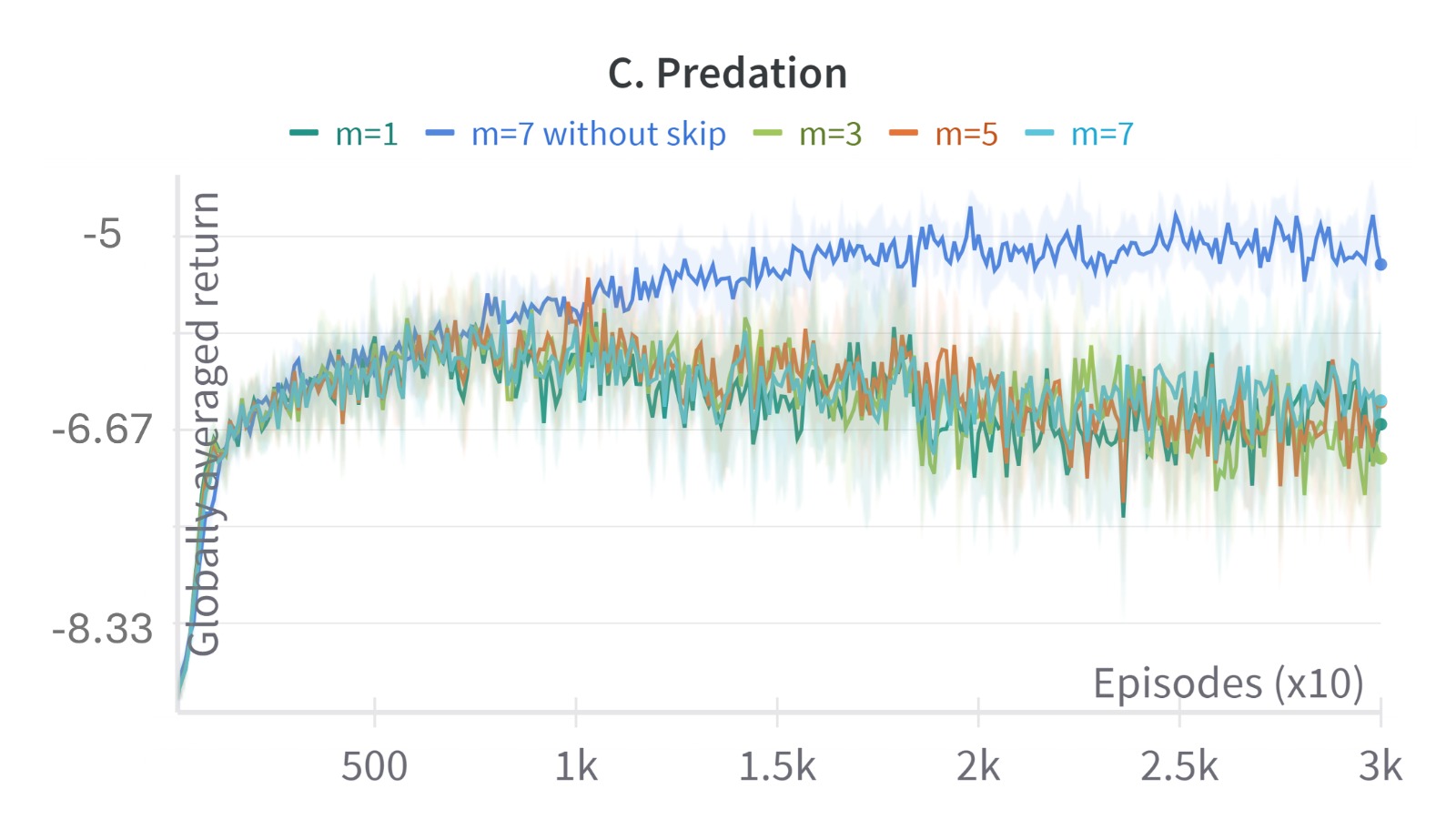}
        \caption{Halving communication frequency.}
    \end{subfigure}
    \caption{Results on C.Predation using skip of communication and neighbours.}
    \label{fig_scalability}
\end{figure}

Fig.~\ref{fig_scalability} shows the results for the C. Predation task with 8 agents. Fig.~\ref{fig_scalability} (a) corresponds to SS with less neighbours, and Fig.~\ref{fig_scalability} (b) with half communication frequency. We compare the results with the default SS algorithm without using the two protocols. The results indicate that reducing the number of neighbours has less influence on the performance than reducing communication frequency. Additionally, compared with the main results shown in Fig.~\ref{main_results_curves} (c), after employing the two protocols to reduce computational costs, SS can still achieve competitive performance.

\subsection{Hyperparameters}\label{hyperparameters}

The hyperparameters used in our experiments are listed in Tables \ref{hyperparameter} and \ref{hyperparameter_}.

\begin{table}[h]
\renewcommand\arraystretch{1.3}
    \caption{Common hyperparameters used in all environments.}
    \centering
    \begin{tabular}{l l l l}
    \toprule [1pt]
     Hyperparameter    &   Value  & Hyperparameter    &   Value\\
     \hline
     Critic learning rate &  1e-4  & Update iteration $K$ &  3 \\
     Discount factor $\gamma$  &  0.99  & Activation & ReLU \\ 
     GAE $\lambda$  &  0.98 &  Optimizer  &  Adam \\
     Clipping $\epsilon$  &  0.2 & & \\
    \bottomrule [1pt]
    \end{tabular}
    \label{hyperparameter}
\end{table}

\begin{table}[h]
\renewcommand\arraystretch{1.3}
    \caption{Hyperparameters used in different environments. }
    \centering
    \begin{tabular}{l l l l l }
    \toprule [1pt]
     Domain    &  Cleanup  &  Harvest & C. Predation  & C. Navigation\\
     \hline
     Critic network size & (1024, 256, 1) & (1024, 256, 1) &  (128, 64, 1)  &  (128, 64, 1) \\
     Actor network size & (1024, 256, d\_a) & (1024, 256, d\_a) & (128, 64, d\_a) &  (128, 64, d\_a)\\
     Actor learning rate & 1e-5 & 5e-5 & 1e-4 & 1e-5\\
     $\rho$  &  1e3  & 0.1  & 0.1 &  1  \\
    \bottomrule [1pt]
    \end{tabular}
    \label{hyperparameter_}
\end{table}

\end{document}